\documentclass[journal]{IEEEtran}
\usepackage{cite}
\usepackage{amsmath,amssymb,amsfonts,bbm,booktabs}
\usepackage{algorithmic}
\usepackage{graphicx}
\usepackage{ragged2e}
\usepackage{textcomp}
\usepackage{xcolor}
\usepackage{lipsum}
\usepackage{fancyhdr}
\def\BibTeX{{\rm B\kern-.05em{\sc i\kern-.025em b}\kern-.08em
    T\kern-.1667em\lower.7ex\hbox{E}\kern-.125emX}}
\usepackage{color}

\newenvironment{mycenter}[1][\topsep]
  {\setlength{\topsep}{#1}\par\kern\topsep\centering}
  {\par\kern\topsep}

\usepackage[caption=false,font=footnotesize]{subfig}

\graphicspath{{figures/}}
\makeatletter
\def\input@path{{./tables/}}
\makeatother

\newcommand{\tcell}[1]{\begin{tabular}[x]{@{}l@{}}#1\end{tabular}}

\newcommand{\Pslr}[0]{P^{\mathrm{Solar}}}
\newcommand{\Qslr}[0]{Q^{\mathrm{Solar}}}

\newcommand{\PFslr}{\mathrm{PF}^{\mathrm{Solar}}}
\newcommand{\cPF}{c^{\mathrm{PF}}}
\newcommand{\PslrMax}[0]{P^{\mathrm{Solar,\,Max}}}

\newcommand{\PslrRate}[0]{\Delta P^{\mathrm{Solar},\,1\%}}

\newcommand{\Psf}{P^{\mathrm{SF}}}
\newcommand{\epsU}{\epsilon^{U}}
\newcommand{\epsI}{\epsilon^{I}}
\newcommand{\epsP}{\epsilon^{P}}

\newcommand{\Vtwin}[0]{U^{\mathrm{DSSE}}}
\newcommand{\Vsend}[0]{U^{\mathrm{SEND}}}

\newcommand{\hPtot}[0]{\hat{P}^{\mathrm{tot}}}

\newcommand{\vsp}[0]{\vspace{0.5em}}

\newcommand{\mrp}{\mathrm{p}}
\newcommand{\mrq}{\mathrm{q}}
\newcommand{\mrpq}{\mathrm{pq}}
\newcommand{\Uline}[0]{U^{\mathrm{LL}}}
\newcommand{\hUline}[0]{\hat{U}^{\mathrm{LL}}}

\renewcommand{\Re}{\mathrm{Re}}
\renewcommand{\Im}{\mathrm{Im}}
\newcommand{\Cbb}[1]{\mathbb{C}^{#1}}

\newcommand{\Urtd}[0]{U^{\mathrm{Rtd.}}}
\newcommand{\Irtd}[0]{I^{\mathrm{Rtd.}}}

\def\BibTeX{{\rm B\kern-.05em{\sc i\kern-.025em b}\kern-.08em
    T\kern-.1667em\lower.7ex\hbox{E}\kern-.125emX}}

\begin{document}

\author{Matthew~Deakin,~Marta~Vanin,~Zhong~Fan,~and~Dirk~Van~Hertem
\thanks{M. Deakin is with Newcastle University, Newcastle-upon-Tyne, UK. M. Vanin and D. Van Hertem are with the electrical engineering department (ESAT) of KU Leuven, Heverlee, Belgium, and with EnergyVille, Genk, Belgium. Z. Fan is presently with the University of Exeter, Exeter, and was previously academic director of SEND at Keele University, UK. E-mail: \texttt{matthew.deakin@newcastle.ac.uk.}}
}

\title{Smart Energy Network Digital Twins:\\Findings from a UK-Based Demonstrator Project}

\maketitle

\begin{abstract}
Digital Twins promise to deliver a step-change in distribution system operations and planning, but there are few real-world examples that explore the challenges of combining imperfect model and measurement data, and then use these as the basis for subsequent analysis. In this work we propose a Digital Twin framework for electrical distribution systems and implement that framework on the Smart Energy Network Demonstrator microgrid in the UK. The data and software implementation are made available open-source, and consist of a network model, power meter measurements, and unbalanced power flow-based algorithms. Measurement and network uncertainties are shown to have a substantial impact on the quality of Digital Twin outputs. The potential benefits of a dynamic export limit and voltage control are estimated using the Digital Twin, using simulated measurements to address data quality challenges, with results showing curtailment for an exemplar day could be reduced by 56\%. Power meter data and a network model are shown to be necessary for developing algorithms that enable decision-making that is robust to real-world uncertainties, with possibilities and challenges of Digital Twin development clearly demonstrated.
\end{abstract}


\section{Introduction}\label{s:introduction}

The digitalization of energy systems continues to gather pace, with new system architectures promising to increase utilization, improve resilience, and support the societal transition towards net zero. These architectures can incorporate real-time monitoring and dispatch in electricity distribution networks, enabling functionalities such as dynamic operating envelopes, flexibility markets, or peer-to-peer energy markets. Digitalization therefore has a central role in transforming passive distribution networks into an active and responsive distribution systems that can host much higher levels of low carbon technologies.

This proliferation of system data is leading both industry and academia to imagine the potential role of Digital Twins in future distribution systems. Whilst definitions of Digital Twins vary substantially between sectors \cite{dalibor2022cross}, for distribution systems they typically combine a network model with measurements to support decision making by network operators and planners. For example, in \cite{gui2023automatic}, the authors propose an online method for assigning Volt/VAr controller set points for low voltage (LV) distribution systems. Similarly, a real time Digital-Twin based method is demonstrated for a multi-energy network in \cite{tian2023digital}. Digital twins have also been proposed for power distribution asset management tasks, such as for substation transformers \cite{moutis2020digital}, or condition monitoring of medium voltage (MV) overhead line systems \cite{gauce2023application}. Industry is also showing a strong appetite for these approaches, with utilities developing increasingly ambitious plans for the development of Digital Twins of their service areas \cite{npg2022digitalization}.

Despite this strong outlook, publications reporting real-life experiences of developing Digital Twins are uncommon, partly due to researchers' limited access to necessary utility data~\cite{MateoDomingo2010}. Typically, a Digital Twin has two steps: a data assimilation stage, reconciling measurement data and system model (e.g., through formulation as a Distribution System State Estimation (DSSE) problem) \cite{Abur2004book, Primadianto}; and an analysis phase, which combines the processed data with other computational methods to support operational or planning decision-making tasks. Despite DSSE having been studied for several decades, academic case studies are typically based on synthetic test cases which fail to show few of the complexities of real networks \cite{GethCIRED2023} and lack the necessary power and voltage measurement data \cite{Lave2019,vanin2023impedance}. For example, a collection of 128 real LV feeders from the UK were developed in \cite{Rigoni2016}, but a lack of appropriate smart meter data meant that simulated active power profiles had to be provided (if DSSE is required, artificial test cases can be created by recording voltages at measurement buses and adding artificial noise \cite{vanin2022framework}). In other cases, power flows and voltages from real networks are made public, but usage of this data can be hampered by a lack of corresponding network model and unspecified measurement accuracy \cite{deakin2023annual}. In contrast, there are many methods for supporting decision making through control- or market-based means \cite{alam2023allocation,zhang2019novel}, but decision-making algorithms typically do not consider the data assimilation process, and so would not nominally be considered a Digital Twin. Such analysis methods may thus fail to fully exploit available information in the decision making process, or make unrealistic assumptions as to data accuracy and robustness. There is therefore a need for test cases which demonstrate practical imperfections in distribution network and measurement data: we show that these are valuable not only for developing `whole' Digital Twin applications, but also for DSSE on its own.

There is a small but growing literature on experiences with real data for use with Digital Twins or DSSE. Several papers report findings and experiences using real data that cannot be shared (e.g., for privacy or commercial reasons). For example,~\cite{zanni2020pmu} present the results of DSSE on a Swiss distribution network, highlighting how mismatches between real and digital line parameters cause their state estimator to flag `normal' current measurements as gross data errors. Groß et al.~\cite{Gross2019} evaluate the sensitivity of DSSE outputs to the pseudomeasurements required in an MV network. Others explore the impact of measurement configuration and error on DSSE results \cite{Frueh2021}, or how incorrect transformer impedances affect the development of Digital Twins \cite{Lave2019}. Finally, a small number of test cases feature both network data and real power data from the network's smart meters, such as the interconnected LV systems in \cite{Koirala2020} and the unbalanced MV/LV test system of \cite{sandell2023}. However, to the authors' knowledge, no prior distribution test cases feature a network model together with both power \emph{and voltage} data. This is because energy (power) data are easier to obtain, as utilities collect them for billing purposes, while voltage measurements are typically not stored. However, errors in the twin's network model can only be identified if voltage measurements are available, and synthetically created voltage measurements do not necessarily capture real-life complexities. Given the role of extensive demand-side electrification in reaching net zero, the need for representative test systems which more closely represent the network and monitoring data available to DNOs in contemporary systems is both significant and timely.

The contribution of this paper is to explore the opportunities and challenges of developing Digital Twins in power distribution systems by proposing a Digital Twin framework for a real-world MV/LV distribution system, then acquiring the models and data alongside a software implementation to realise this twin. Opportunities to provide inputs for decision-making for system owners are demonstrated as a clear potential benefit of the Digital Twin, whilst also highlighting the risks if an insufficient model and data fidelity cannot be reconciled to create a consistent system representation. The network model \cite{Deakin2023sendNtwk} and measurement data \cite{Deakin2023sendMsmnt} are made available open-source, with a workbook for reproducing results of this work at: 

\begin{mycenter}[0.2em]
\texttt{https://github.com/deakinmt/DSSE\_SEND}.
\end{mycenter}

The paper is structured as follows. Section~\ref{s:framework} proposes a Digital Twin for power distribution networks, highlighting the crucial requirements for DSSE within the Digital Twin for data assimilation. Section~\ref{s:send} provides a detailed explanation of the Smart Energy Network Demonstrator (SEND) microgrid, on which the Digital Twin framework is implemented. Section~\ref{s:towards_digital_twin} discusses the challenges encountered working with the real data, showing how the present meter configuration and network model require improvement to reach the standards needed for effective data assimilation. The benefits of a dynamic voltage control scheme are estimated in Section~\ref{s:curtailment_cases} to highlight a potential use-case of the Digital Twin, albeit using the Digital Twin in a synthetic setup (i.e., using a combination of real and simulated `measurements'). In Section~\ref{s:next_test_case} we survey potential future directions and research gaps for Digital Twins, before offering conclusions in Section~\ref{s:conclusions}.

\section{A Digital Twin Framework for Distribution Network Applications}\label{s:framework}

\begin{figure}
\centering
\includegraphics[width=0.48\textwidth]{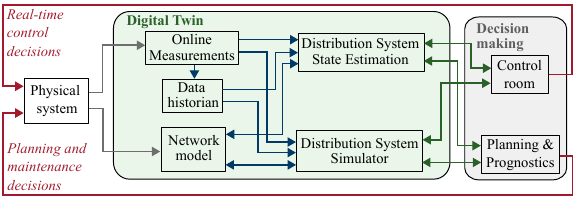}
\caption{The proposed Digital Twin framework aims to combine network models and telemetry with DSSE and simulation algorithms. Section~\ref{s:send} discusses the physical system, network model, online measurements, and data historian; Section~\ref{s:towards_digital_twin} discusses the network model and DSSE; Section~\ref{s:curtailment_cases} discusses distribution system simulation and decision making activities.}
\label{f:digital_twin_workflow}
\end{figure}

Definitions of Digital Twins vary significantly between sectors and applications \cite{dalibor2022cross}. For the purposes of this work, we consider a distribution system Digital Twin to be characterized by five elements that together represent the physical system, as illustrated in Fig.~\ref{f:digital_twin_workflow}. Functionally, the Digital Twin acts as an interface for supporting decision making in both planning and operational decision making processes.

The Digital Twin takes input data from the physical system in the form of a network model and online measurements. The network model includes all information required to run a power flow, i.e., network topology, impedances (cables types and lengths), transformer models, and load/generator connectivity. Measurement data from power meters and other data acquisition systems can be used online and stored in a central data historian for subsequent interrogation. The network model, online measurements and data historian can be used for simulation of the distribution system in unrealized circumstances (using a distribution system simulator), or for estimating the state of the system given noisy measurements (using DSSE algorithms). Table~\ref{t:digital_twin_tools} lists the software used to carry out these functionalities.

\begin{table}
\caption{Tools used to realise the SEND Digital Twin.}
\centering
\begin{tabular}{ll}
\toprule
Digital Twin Module & Tool or format \\
\midrule
Online measurements & Siemens SICAM P855 \vsp\\
Data historian & \tcell{Siemens Distributed Energy\\Optimization (DEOP);\\cached in \texttt{.csv} format}\vsp\\
Network Model & OpenDSS \texttt{.dss} format \vsp \\
\tcell{Distribution System\\Simulator} & \tcell{OpenDSS (nonlinear power\\flow); Julia (linear\\power flow)} \vsp \\
\tcell{Distribution System\\State Estimation\\(DSSE)} & \tcell{PowerModelsDistribution\\StateEstimation.jl\\ \cite{vanin2022framework}} \\
\bottomrule
\end{tabular}

\label{t:digital_twin_tools}
\end{table}

\subsection{Comparing Digital Twins with Existing Practise}

In transmission networks, data acquisition and assimilation are part of state estimation workflows since the 1970s, and control rooms receive updated information on the system's (steady) state every few minutes. Transmission networks are in general fully observable, and measurement semantics are well understood. Therefore, Digital Twins could be seen as an extension of current control room practise, rather than the development of a new operating approach.

In contrast, distribution network practices vary hugely depending on geography, regulation, and policies. It is thus challenging to provide a clear overview of the state of the art and compare it with the proposed Digital Twin. However, in many distribution systems, developing a Digital Twin would represent a substantial change in the utility operating practise.

Typical challenges that distribution network Digital Twins need to address include integrating disparate databases (e.g., combining GIS information directly with parameters such as line construction codes); providing seamless interfaces to third-party information (e.g., AMI data) as and when required; and improved information recording for critical parameters such as switch states or off-load transformer tap positions. Additional organizational bottlenecks and more informed engineering practices are discussed in~\cite{GethCIRED2023}, such as enhanced measurement semantics, better integration of data sources, and systematic methods for validating electrical network models.

\section{The Smart Energy Network Demonstrator}\label{s:send}

\begin{figure}
\centering
\includegraphics[width=0.48\textwidth]{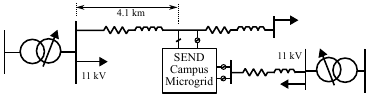}
\caption{The SEND virtual private wire system (or `microgrid') interconnects with the wider distribution network at four points.}
\label{f:mv_system}
\end{figure}

The Smart Energy Network Demonstrator (SEND) \cite{fan2022role} is located at the Keele University campus in the West Midlands of the UK, and aims to provide an energy system `living laboratory' \cite{fan2022role}. From the distribution system perspective, this includes substantial investment in power quality metering equipment and the associated data collection platform. As shown in Fig.~\ref{f:mv_system}, the SEND electrical system operates as a virtual private wire system, i.e., it is fully owned and operated by the university, rather than by the local network operator. As a result, the university is responsible for all network and measurement data. The system is therefore an ideal sandbox to explore challenges and opportunities for Digital Twins. This section presents the network model and measurement data for SEND that were used to implement the Digital Twin framework (as described in Section~\ref{s:framework}).

\subsection{Network Topology and Model}

\begin{figure}
\centering
\includegraphics[width=0.44\textwidth]{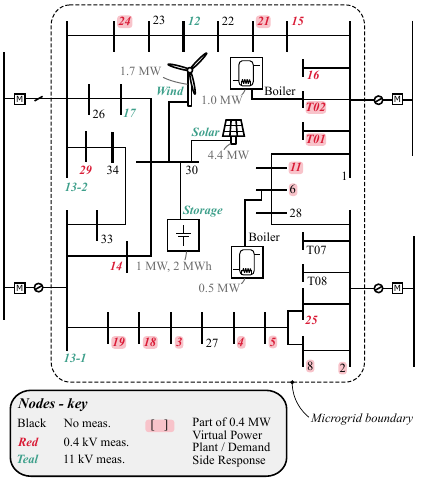}
\caption{Augmented single line diagram of the MV network of the SEND virtual private wire network showing the network, distributed energy resources and location and voltage level of power meters. For conciseness, LV substation transformers are not shown.}
\label{f:send_sld}
\end{figure}

The augmented single line diagram in Fig.~\ref{f:send_sld} shows the electrical network topology, the location of distributed energy resources, and the voltage and location of the power meters. A summary of key network parameters is reported in Table~\ref{t:network_summary}, showing that the network peak generation is much higher than the baseload demand. The topology of the MV network is weakly meshed, with LV substations (not shown) feeding demands connected radially. Within the LV networks, only the LV substation busbar is modelled.

\begin{table}
\caption{Summary of SEND microgrid parameters.}
\centering
\begin{tabular}{ll}
\toprule
Parameter & Value\\
\midrule
Number of MV buses & 33 \\
Total no. nodes & 174 \\ 
Export limit (buses 13-1, 13-2) & 1.578~MVA\\ 
Export limit (buses 1, 2) & 0~MVA\\
Total AC generation capacity & 7.1~MVA\\
Minimum Demand & 0.24~MVA\\ 
Baseload Demand & 4.03~MVA\\ 
Total MV cable length & 13.1~km\\ 
MV upper voltage limit & 1.06~pu\\
LV upper voltage limit & 1.10~pu\\
MV lower voltage limit & 0.94~pu\\
LV lower voltage limit & 0.94~pu\\
\bottomrule
\end{tabular}

\label{t:network_summary}
\end{table}

There are four potential points of coupling to the upstream network (as also shown in Fig.~\ref{f:mv_system}): bus 1, 2, 13-1, or 13-2. The network is normally operated with only bus 13-2 connected to the main distribution network, as this is the point of common coupling (PCC) for which the microgrid can export. However, even from this point of connection, voltage constraints mean that the export at the point of common coupling must remain less than 1.578~MW. As the total connected generation capacity is greater than this value, the site is subject to an export limit, with protection in-place so that the generators will trip if there is a greater export than this value for a period of more than 10 seconds \cite{ena2023g100}. To avoid tripping of the protection, an alarm is therefore raised after one second by the energy management system, then after a further four seconds generation is curtailed by the energy management system. It is therefore not uncommon to see oscillatory, `hunting' behaviour in PV output during high solar output, as the generation is reduced due to over-export, slowly ramped up over the course of a few minutes, and then reduced once again due to the over-export (as considered in more detail in Section~\ref{s:curtailment_cases}).

\subsubsection{Distributed Energy Resources}\label{sss:ders}

Fig.~\ref{f:send_sld} shows that in addition to the generators, there are a number of other distributed energy resources (DERs). Together, these DERs can be scheduled to reduce both the total cost of energy supplied to the SEND campus and overall campus carbon intensity. For example, when surplus generation is predicted, the site's gas boilers can be turned off and the campus water heating demand is instead met via three 0.5~MW electric boilers.

In addition to the boilers, approximately 400 kVA of demand side response technologies are split across a wide range of different demand classes, summarised in Table~\ref{t:vpp_summary}. It is interesting to note that, even though there are more than 150 assets, together they provide less flexibility than just a single large electric boiler. Nevertheless, they still represent a significant opportunity to shift demand temporally to make use of renewable energy that would otherwise be spilled.

\begin{table}
\caption{Virtual Power Plant assets, together representing approximately 400 kVA demand-side response.}
\centering
\begin{tabular}{ll}
\toprule
Parameter & Number \\
\midrule
Air conditioning unit & 1 \\
Air handling units & 24 \\
Pump & 5 double, 11 single\\
Fan coil units & 20 \\
Electric boiler & 3 \\
Smart plug & 95 \\
Immersion heater & 10 \\
Chiller & 7 \\
Car Chargers & 11 dual chargers \vspace{0.3em} \\
\tcell{Air Handling Unit\\Circuit Pumps} & 1 double \vspace{0.3em} \\
\tcell{Variable Temperature\\Circuit Pumps} & 1 double \\
\bottomrule
\end{tabular}

\label{t:vpp_summary}
\end{table}

\subsection{Telemetry}\label{ss:power_meters}

As shown in Fig.~\ref{f:send_sld}, power meters are spread throughout the network. Table~\ref{t:telemetry_summary} summarises their settings. At present, there are twenty two monitoring devices connected to the distribution network (as SEND continues to develop, this number is growing). These meters can measure a wide range of relevant power quality phenomena \cite{siemens2020sicam}, and are configured to record rms line voltages, phase currents, three-phase real and reactive power, each at a 30 second interval (Table~\ref{t:network_summary}). The recorded data is transmitted to a data collection platform (Siemens' Distributed Energy Optimization Platform, DEOP), where data can be visualized or queried from an API for further analysis. Eight days of data have been scraped from the API and provided in \texttt{csv} format. These dates represent the second Friday and Saturday of the month every other month from May to November 2022, and are given in Table~\ref{t:send_dates}. These dates cover weekday and weekends, and represent between them periods of low, moderate and high system demand (undergraduate students are vacant from the campus in June and July).

\begin{table}
\caption{Dates in 2022 of data scraped from the SEND DEOP data historian API.}
\centering
\begin{tabular}{ll|ll}
\toprule
Date & Weekday & Date & Weekday\\
\midrule
13th May & Friday & 16th Sept. & Friday\\
14th May & Saturday & 17th Sept. & Saturday\\
15th July & Friday & 11th Nov. & Friday\\
16th July & Saturday & 12th Nov. & Saturday\\
\bottomrule
\end{tabular}

\label{t:send_dates}
\end{table}

\begin{table}
\caption{Summary of SEND electrical telemetry parameters.}
\centering
\begin{tabular}{ll}
\toprule
Parameter & Value\\
\midrule
Meter type & \tcell{Siemens SICAM P855 \cite{siemens2020sicam}} \\
No. MV meters & 7\\ 
No. LV meters & 15\\
Synchronization accuracy & $ \pm $ 25ms \\
Update frequency & Twice per minute \vsp \vsp\\
Measurands & \tcell{Line-line voltages\\magnitudes [V], phase\\currents [A], apparent\\power [kVA], total\\real power [kW], total\\reactive power [kVAr],\\ power factor.}\\
\bottomrule
\end{tabular}

\label{t:telemetry_summary}
\end{table}

\begin{figure}\centering
\subfloat[Voltages]{\includegraphics[width=0.16\textwidth]{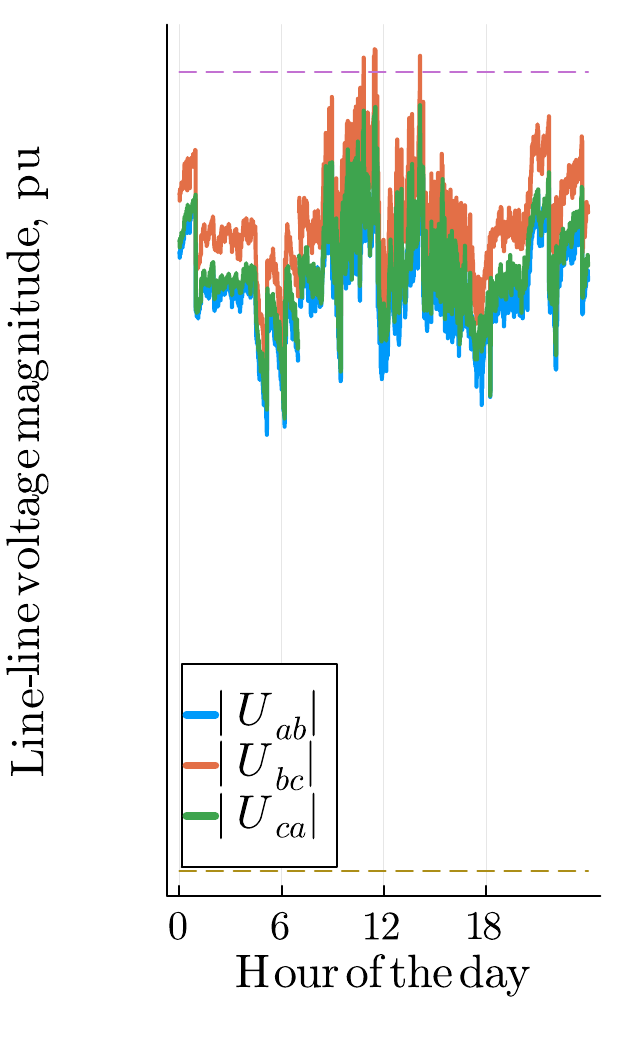}\label{f:pltSendSolar_v}}
~
\subfloat[Currents]{\includegraphics[width=0.16\textwidth]{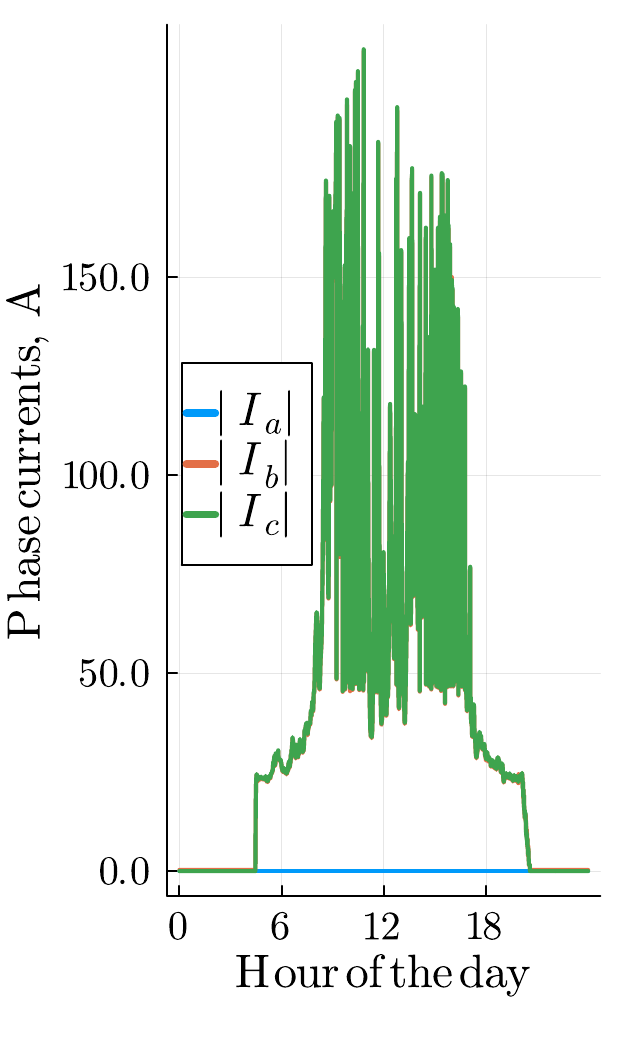}\label{f:pltSendSolar_i}}
~
\subfloat[Powers]{\includegraphics[width=0.16\textwidth]{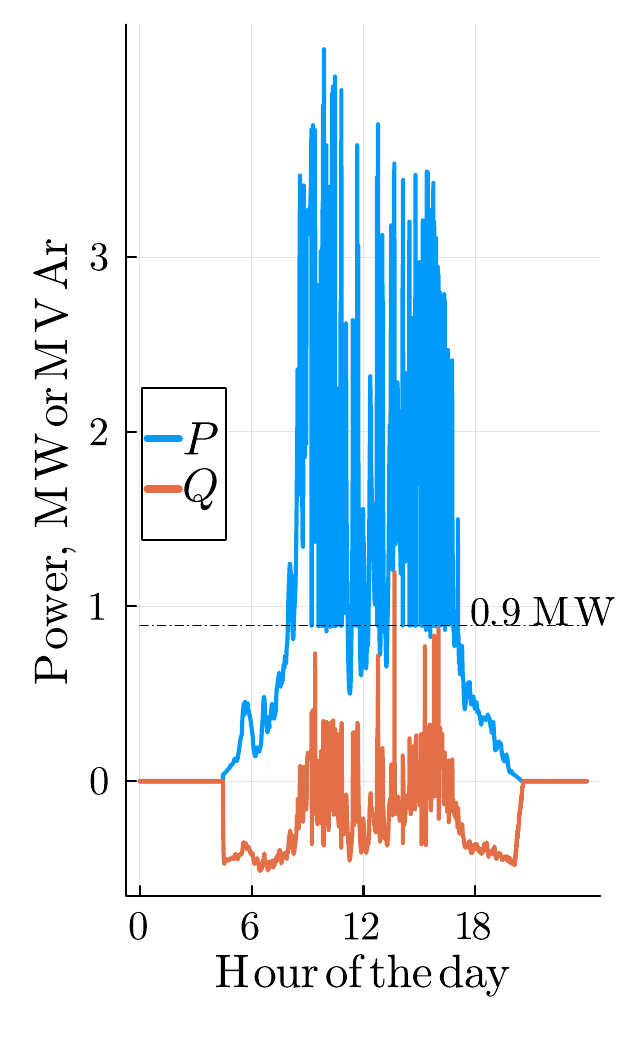}\label{f:pltSendSolar_s}}
\caption{Voltages, currents and power measurements for the solar PV generator connected close to bus 30 for 15th July 2022.}
\label{f:pltSendSolar}
\end{figure}

Fig.~\ref{f:pltSendSolar} shows the measured voltages, currents and powers for the solar bus on 15th July 2022. The line-line voltages are high, and exceed the nominal MV voltage limit of 1.06~pu for short periods of time. Currents and powers, as expected, increase through the morning and reduce through the evening, although the measured phase-$a$ current is consistently zero (this and other data quality issues are addressed in Section~\ref{ss:data_quality}). On this date, there is also solar curtailment, as can be observed in the power injection shown in Fig.~\ref{f:pltSendSolar_s} (on this day, if the export limit is crossed, the solar PV output is reduced to 0.9 MW).

\begin{figure}
\centering
\includegraphics[width=0.48\textwidth]{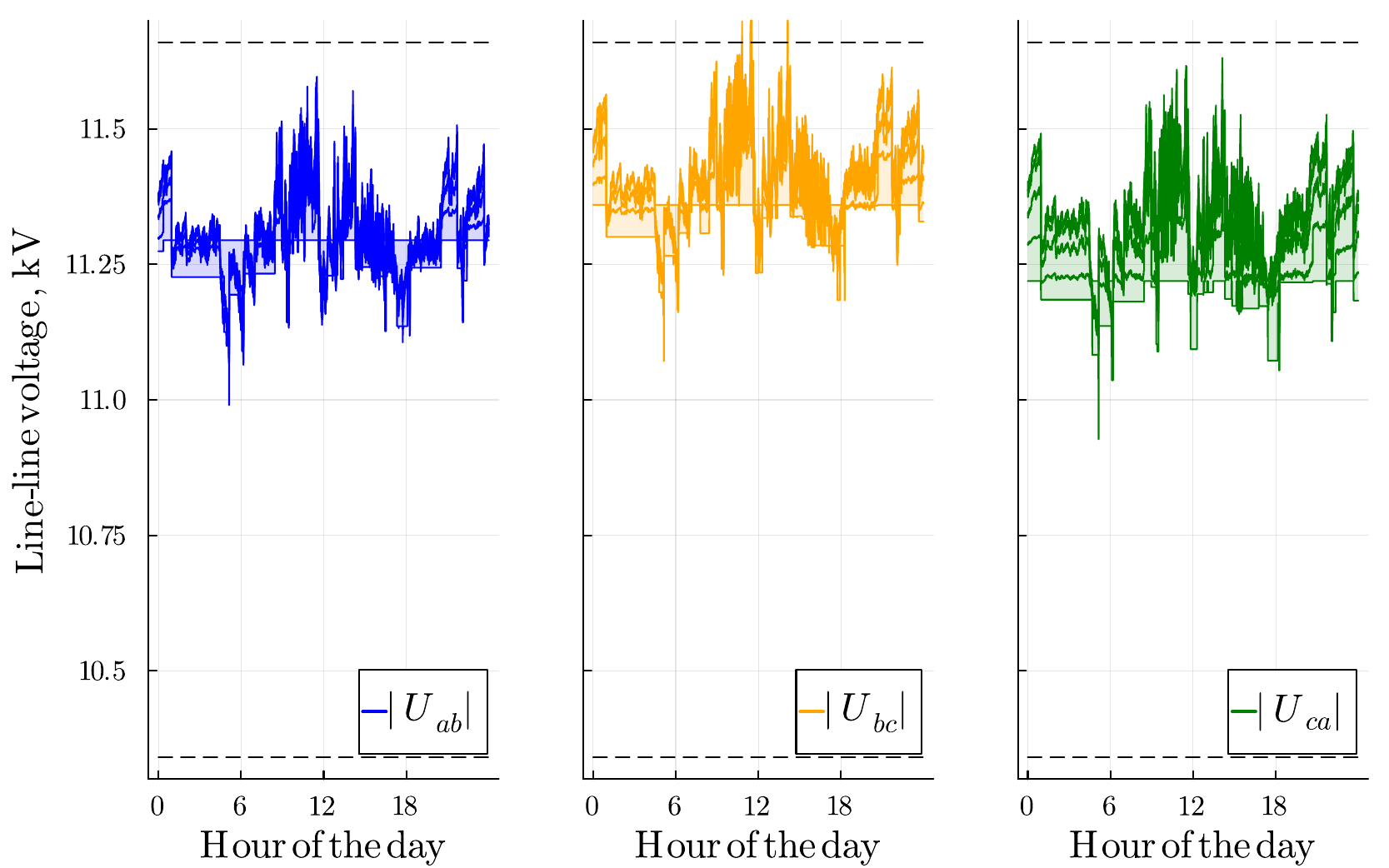}
\caption{Range, interquartile range and median MV line-line voltage measurements across all buses for 15th July 2022.}
\label{f:pltSendMvVoltages}
\end{figure}

\begin{figure}
\centering
\includegraphics[width=0.48\textwidth]{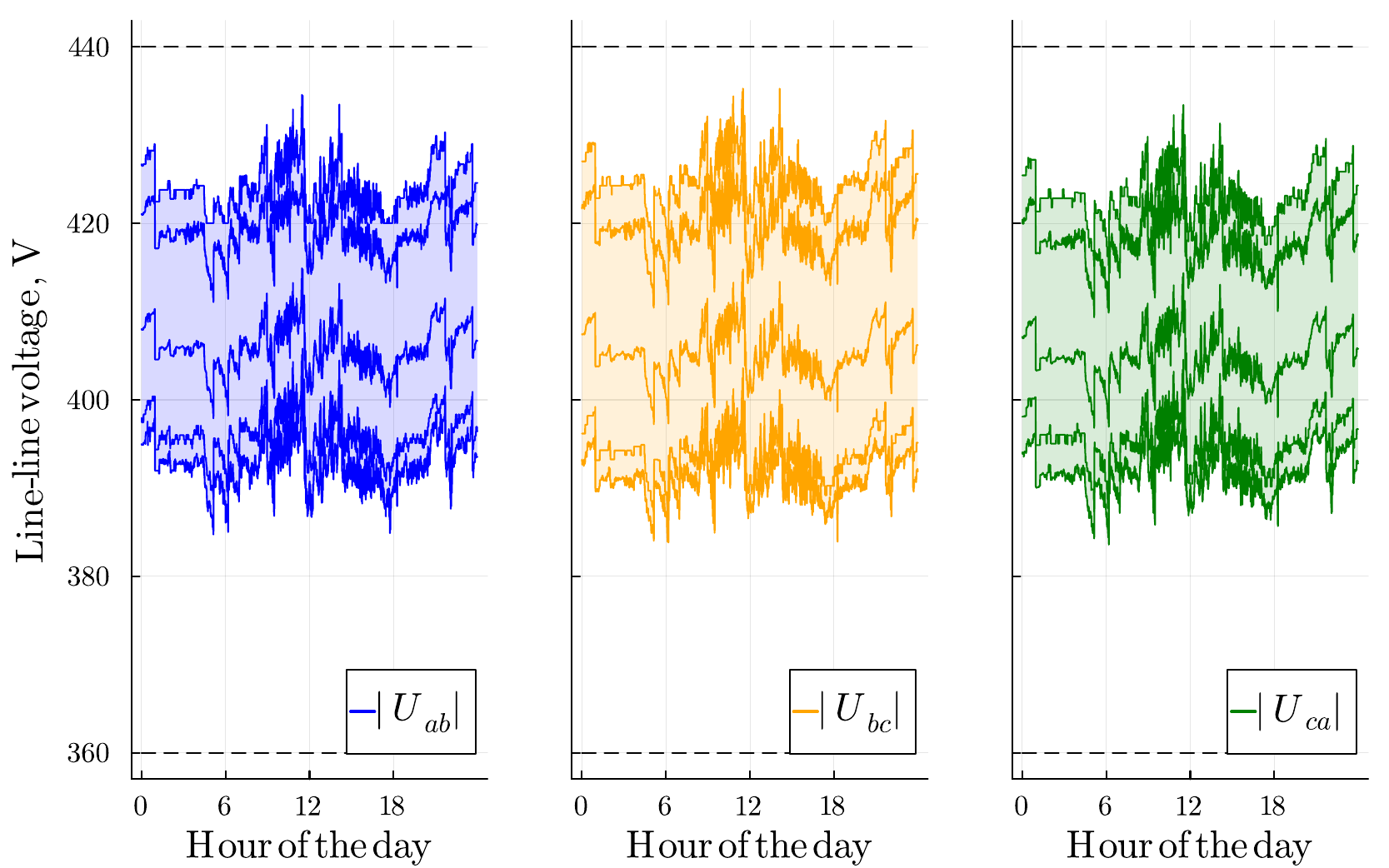}
\caption{Range, interquartile range and median LV line-line voltage measurements across all buses for 15th July 2022.}
\label{f:pltSendLvVoltages}
\end{figure}

Naturally, voltages also change across the network through the day. This can be visualised by considering how the range, interquartile range and median voltage varies, as plotted in Fig.~\ref{f:pltSendMvVoltages} and Fig.~\ref{f:pltSendLvVoltages} for the MV and LV networks. By comparison with Fig.~\ref{f:pltSendSolar_v}, it can be observed that over the course of the day that voltage magnitudes follow the pattern of solar voltages, but that the spread is much greater for the LV voltages. This increase in spread is due to different turns ratios for the transformers in the network (discussed in Section~\ref{s:towards_digital_twin}).

\subsubsection{Modelling Measurement Uncertainty}\label{sss:tolerances}

The measurement uncertainty and valid measurement range of the power meters vary for different measurands, as shown in Table~\ref{t:relative_errors_summary}. Assuming the meter's rated voltage $\Urtd $ is the equivalent to the per-unit operating voltage of the network, these meters will capture voltages during normal operations with operational uncertainty $\epsU$ of 0.5\% (voltage bounds on MV networks in the UK are between 0.94 and 1.06~pu \cite{uk2002escqr}, and so during normal operation network voltages are less than 120\% of $\Urtd$). 

\begin{table}
\caption{Measurement relative errors for the SICAM P855 (from the datasheet \cite{siemens2020sicam}).}
\centering
\begin{tabular}{lll}
\toprule
Measurand & \tcell{Operational\\Uncertainty} & Valid Range\\
\midrule
\tcell{Line-line\\voltage} & 0.5\% & 0\% to 120\% of $ \Urtd $ \\
\tcell{Phase\\currents} & 0.2\% &  20\% to 200\% of $ \Irtd $  \\
\tcell{Real\\Power} & 1\% &  20\% to 200\% of $ \Irtd $ \\
\tcell{Reactive\\Power} & 2\% & 20\% to 200\% of $ \Irtd $ \\
\bottomrule
\end{tabular}

\label{t:relative_errors_summary}
\end{table}

Current and power measurements have relative error of 0.2\% for currents between 20\% and 200\% of the meter's rated current $\Irtd$. These flows are often much less than 20\% of $\Irtd $, and so the accuracy is not specified at those ratings. Here, it is assumed that the absolute current measurement error $\epsI_{i,\,\phi}$ does not deteriorate as the current reduces below this value (subscripts indicate indexing, with $i$ the bus index and $\phi$ the phase index). That is, for the current measurement on bus $i$ and phase $\phi$, $\hat{|I|}_{i,\,\phi}$, the absolute current operational uncertainty does not drop below the value at 20\% of the rated current for the device at bus $i$, $ \Irtd_i $, i.e.,
\begin{equation}\label{e:tolerance_epsI}
\epsI_{i,\,\phi} = 0.2\% \times \max \left \{ \hat{|I|}_{i,\,\phi},\,0.2\times \Irtd_i \right \} \,.
\end{equation}
We note that in academic DSSE papers, the valid region of the operational uncertainty is typically neglected, leading to very optimistic assumptions as to current and power measurement uncertainty at loading below 20\% as compared to \eqref{e:tolerance_epsI}.

\section{Assessing Digital Twin Adequacy via Distribution System State Estimation}\label{s:towards_digital_twin}

For a Digital Twin to fulfil its role as a decision-making tool, measurement data and network models need to be assimilated to create a meaningful, non-trivial and self-consistent system representation. In the context of the SEND Digital Twin, the voltage outputs of DSSE and distribution system simulations (power flow) can be compared to historic measurement data to perform this validation. A functioning Digital Twin will have DSSE outputs that are within tolerance of meter operational uncertainties, with large differences between measured and calculated variables being associated with low confidence in the Digital Twin's performance. In the case of inadequate performance, Digital Twins can still be studied as synthetic, `virtual' Digital Twins (considered in Section~\ref{s:curtailment_cases}).

In this section, we show that the data quality and network modelling accuracy in the present configuration need to be improved for the real-world system at SEND to achieve the high standard required to be an effective Digital Twin. Firstly, Section~\ref{ss:data_quality} discusses the quality of raw data from power meters, showing how a number of data quality issues require preprocessing. Challenges in network model development are outlined in Section~\ref{ss:tap_settings}, considering data inconsistency and incomplete component information. In particular, inaccurate MV/LV transformer models are shown to have a significant detrimental impact on the Digital Twin's performance. Finally, challenges around power meter configurations are outlined in Section~\ref{ss:dsse_and_observability}. We show that, even assuming a perfect network model and noiseless measurements, meter paucity and inadequate measurement semantics make it impossible to recover the state of the system reliably. A bespoke, heuristic DSSE approach is proposed to address these issues, but a clear need for improved data quality is seen (with regards to measurements and network model) to enable real-world DSSE-based decision making.

\subsection{Measurement Data Quality}\label{ss:data_quality}

The power meters installed in SEND (Section~\ref{ss:power_meters}) are designed to collect 30 second rms measurements continuously to a prespecified measurement accuracy (Table~\ref{t:relative_errors_summary}). However, by inspection of the recorded data, it is clear that the measured data does not meet that accuracy in all cases.

\begin{figure*}\centering
\subfloat[Fixed measurements]{\includegraphics[width=0.325\textwidth]{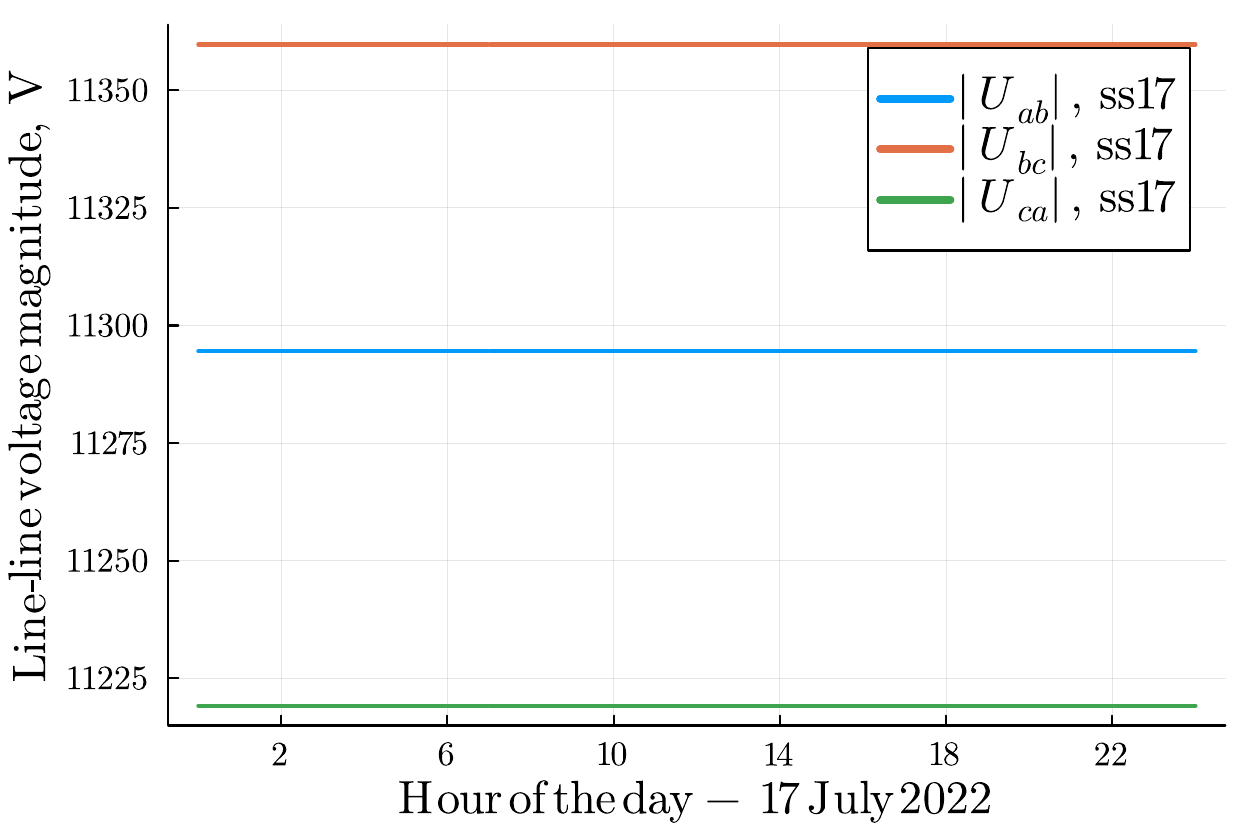}\label{f:data_quality1}}
\subfloat[Threshold configuration and fixed measurement]{\includegraphics[width=0.325\textwidth]{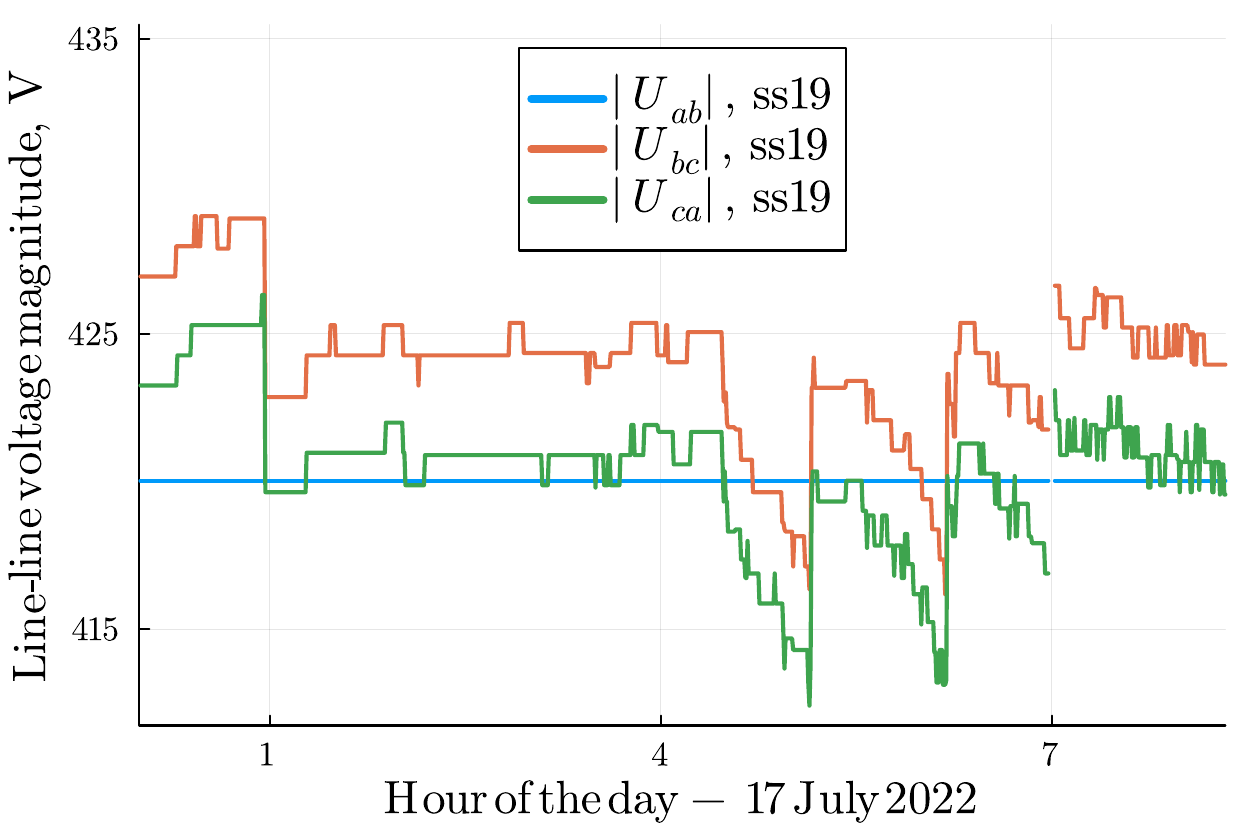}\label{f:data_quality2}}
\subfloat[Gross error]{\includegraphics[width=0.325\textwidth]{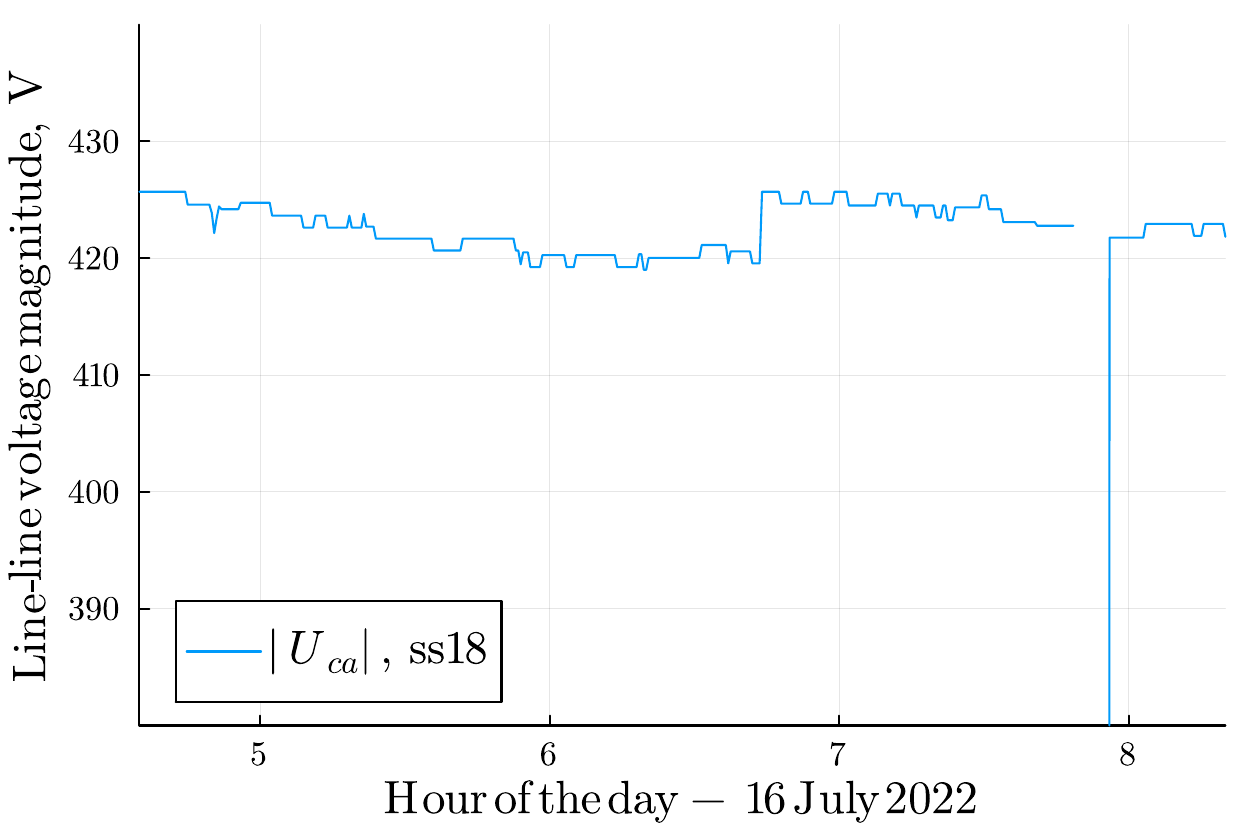}\label{f:data_quality3}}
\caption{Three types of errors identified by inspection.}
\label{f:data_quality}
\end{figure*}

Some data quality issues can be observed in the raw plotted data in Fig.~\ref{f:data_quality}. In some cases, measurement errors are trivial, such as meters that are `stuck' on a single value for each measurement, like that in Fig.~\ref{f:data_quality1} (an issue also seen for $I_{\mathrm{a}}$ in Fig.~\ref{f:pltSendSolar_i}). Other meters show mixed problems, such as in Fig.~\ref{f:data_quality2}: one of the measurement values is fixed, whereas the two other measurements follow a stepped pattern. This occurs if the device only records a change if some threshold is met (for example when the measurand has changed by some fixed percentage). Furthermore, individual time stamps may present a trivial gross error or missing value, as shown in Fig.~\ref{f:data_quality3}. Finally, a small number of meters have been installed but are yet to be configured, and so return non-numerical values (NaNs). It is worth noting that new meters returning large errors is often due to incorrect firmware settings (firmware can typically be updated rather than requiring a new meter to be installed).

Finally, some phenomena that might be associated with data quality issues were identified as measurement tolerance issues. An example is that of the solar plant meter measuring small power demands overnight (not plotted). Such event could suggest that a load is connected but ended unreported in the network data. However, as overnight currents are $< 20\%$ of the rated one, the solar meter presents a large relative error (see \eqref{e:tolerance_epsI}), which leads to powers with an incorrect sense (i.e., `wrong sign' for the active power). It is interesting to note that sense for the power that is consistently incorrect indicates bias in the measurements, contrasting with the common assumption of zero-mean Gaussian measurement error (which would imply both positive and negative errors during times when there is no generator export).

\subsection{Network Model Development}\label{ss:tap_settings}

The Digital Twin's network model was built considering the type and length of MV cables connecting the substations, with sequence components being used to create cable impedances. In general, the data for creating this model (obtained in a previous project) was not fully consistent, and some elements (system topology, cable types) required confirmation either through review of GIS information or via discussion with university estates. While we assume their impact to be modest, potential inaccuracies in those elements' values remain.

A more substantial challenge rests in the estimates of the tap positions of the MV/LV transformer models. The transformers turns ratios can influence the output voltage by double-digit percentages across their full range, and the incorrect choice of tap positions in the Digital Twin may lead to substantial simulation and estimation errors. Transformer models are based on provided datasheets that report winding types, impedances, and nominal voltages for primary and secondary windings. However, it is known that manual tap positions changes may go unreported~\cite{GethCIRED2023}, and so actual tap positions may be very different from their nominal positions.

\begin{figure}\centering
\subfloat[Nominal tap positions]{\includegraphics[height=4.4cm, trim = {0cm, 2cm, 10cm, 2cm}, clip]{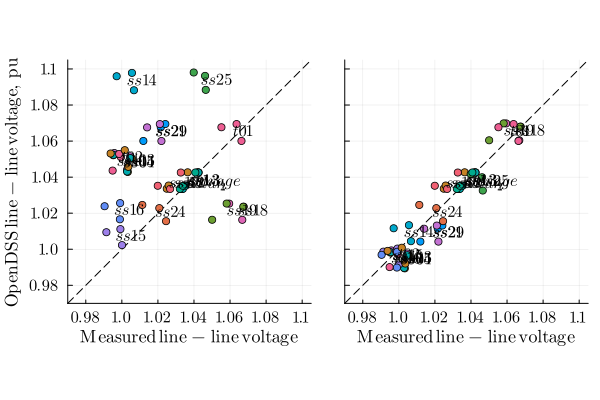}\label{f:pltScatterPre}}
\subfloat[Updated tap positions]{\includegraphics[height=4.4cm, trim = {11.5cm, 2cm, 0cm, 2cm}, clip]{pltScatterTaps_jl.png}\label{f:pltScatterPost}}
\caption{Comparing simulated versus measured line-line voltages, for 12.14pm on September 17th, before and after updating MV/LV tap positions. Each color represents three measurements at a single measurement point. Updated taps lead to a much closer fit (the dashed line represents the line $ y=x $, a perfect fit would lie exactly on this line).}
\label{f:pltScatterVoltages}
\end{figure}

To consider the accuracy of taps as nominally provided, a non-linear unbalanced power flow was run, with the SEND network model's PCC (ss13-1) serving as the slack bus. To convert from line to phase voltages, slack bus voltage angles are assumed 120$^{\circ }$ apart and then magnitudes determined from the measured line voltages. Loads are assigned according to their measured power or according to a nominal split of unaccounted power at the PCC. The power flow solution with nominal tap positions is shown in Fig.~\ref{f:pltScatterPre}, with each of the three points corresponding to the meter's three line voltages. Large discrepancies can be observed between measured and simulated voltages: for example, simulated voltages at `ss14'\footnote{Substations are referred to according to their reference in the network model \cite{Deakin2023sendNtwk}, as compared to the simpler single line diagram of Fig.~\ref{f:send_sld}.} are much higher than measured values, whilst those of `ss18' are lower. While not plotted, the pattern of Fig.~\ref{f:pltScatterPre} is consistent over time, and is largely independent of assumptions on network loading.

It is possible to reduce this error via a trial-and-error heuristic by manually increasing or decreasing the MV/LV tap positions to yield a reasonable fit, with the improvement in fit shown in Fig.~\ref{f:pltScatterPost}. This resulted in the simulated and measured line voltages becoming much more aligned. In principle, a field inspection could confirm the transformer type and position at all locations; however, this is resource-intensive. We note that self-correcting network models are a desirable feature for future-proof Digital Twins~\cite{GethCIRED2023}. However, due to meter paucity, the corrections could not be automated, i.e., a human-in-the-loop was required for this exercise. `Augmented' DSSE examples exist, with tap ratios included as problem variables alongside conventional states (e.g.,~\cite{Nanchian2017}). These augmented DSSE require full observability, a property which does not hold in SEND (as described in Section~\ref{ss:dsse_and_observability}).

\subsection{DSSE and Observability}\label{ss:dsse_and_observability}

In general, reliable DSSE requires both a very good network model and measurement devices which are appropriately placed and configured. With regards to the latter, there are three specific challenges in the SEND measurement system today. Firstly, power meters are configured to report line voltages $\Uline$ rather than phase voltages; line voltages do not have a common reference or neutral, and so zero sequence voltages cannot be detected. As phase voltages are typically considered to be the output state of DSSE, unbalanced DSSE algorithms usually assume the availability of phase voltage measurements. DSSE implementations must therefore be adjusted to take line voltages as input (but nonetheless without a reference voltage, will be unable to uniquely specify voltages with respect to ground).

Secondly, the meters are configured to report only total power flows, rather than per-phase powers. Per-phase power measurements are particularly useful for DSSE as they implicitly include information on the phase angle between local current and voltage phasors, and the current angle has a significant impact on voltage drops across a multi-phase impedance. Furthermore, as the DSSE state typically includes per-phase power variables, providing only their sum (one measurement per three-phase load) reduces observability as compared to a measurement strategy utilizing three per-phase measurements per load. Therefore, contemporary unbalanced DSSE algorithms are typically reliant on per-phase power measurements \cite{vanin2022framework}.

Lastly, the power meters are only placed in a limited subset of nodes, resulting in an unobservable system. In contrast to power meter configuration issues, this scarcity of measurement devices is a well-known challenge for DSSE \cite{Primadianto}.

In principle, all three of the above issues could be addressed. Power meters can be configured to provide per-phase voltages and powers, and in some situations utilities are requesting this as a requirement for the meters that are being installed (e.g., \cite{npg2022technical}). Power meters can also be retrofitted to existing substations to improve observability (note that in SEND, no measurement devices are available other than at the substations highlighted in Fig.~\ref{f:mv_system}). However, in this instance, it proved to be too costly to increase the number of power meters or reconfigure the meters and data acquisition system. Furthermore, whilst there may be a very wide set of potential meter configurations (e.g., choice of measurands, accuracy ratings, percentage substation coverage) which can support DSSE from a theoretical perspective, it may only be viable for commercial DSSE solutions to use the most reliable and well-understood subset of those meter configurations. Adjusting meter settings and software implementations post-deployment may be uneconomical, and so utilities and practitioners should strive to specify measurement requirements adequately prior to meter, firmware and DSSE deployment.

\subsubsection{DSSE Approach and Results}\label{sss:dsse_approach}
To run DSSE, we use the open-source PowerModelsDistributionStateEstimation.jl package \cite{vanin2022framework}. This tool is attractive for the SEND system as it can solve under-determined state estimation problems without relying on pseudomeasurements, and because the optimization-based framework is easily customisable, enabling bespoke measurement constraints such those necessary to include line voltage measurements$ \Uline $. This ease-of-prototyping is possible thanks to the underlying automatic differentiation toolbox JuMP \cite{JuMP}. 

As is conventional for DSSE problems, measurement errors are assumed to be independent and normally distributed, with standard deviation one third of the maximum error,
\begin{equation}\label{e:voltage_distribution}
\Uline _{i,\,\mrpq} \sim \mathcal{N} \left ( \hUline_{i,\,\mrpq};\,\dfrac{0.005 \times \hUline_{i,\,\mrpq}}{3} \right )\,,
\end{equation}
where $ \hUline_{i,\,\mrpq} $ is the line voltage measurement at bus $i$ measured from phase $\mrp$ to phase $\mrq$.

Without per-phase current or power flow measurement, the system would be unobservable even if all nodes had phase voltage and three-phase power measurements. Therefore, for the purposes of this work, we combine current magnitudes $|\hat{I}|_{i,\phi}$ and total powers $\hPtot_i$ into a heuristic composite per-phase power measurement $\hat{P}_{i,\,\phi}$ for phase $\phi$ for the $i$th bus,
\begin{equation}\label{eq:power_per_phase_approx}
\hat{P}_{i,\,\phi} = \hPtot_i \dfrac{ |\hat{I}|  _{i,\phi } }{ \sum_{\phi \in   \{a, b, c \}} |\hat{I}|_{i,\phi }} \,.
\end{equation}
A similar identity holds for a per-phase reactive power. The uncertainty of \eqref{eq:power_per_phase_approx} is not known, but is required for DSSE. This uncertainty is based on \eqref{e:tolerance_epsI} and is estimated according to
\begin{equation}\label{e:tolerance_epsPph}
\epsP_{i,\phi } = 0.01 \times \Urtd_i \times \max \left \{ \hat{|I|} _{i,\phi },\,0.2\times \Irtd_i \right \} \,,
\end{equation}
leading per-phase powers to be distributed as
\begin{equation}\label{e:phase_power_distribution}
P_{i,\phi } \sim \mathcal{N} \left ( \hat{P}_{i,\phi };\, \dfrac{\epsP_{i,\phi }}{3} \right )\,.
\end{equation}
An equivalent distribution exists for per-phase reactive power measurements (with the exception that the operational tolerance of 1\% is replaced with 2\%, as in Table~\ref{t:relative_errors_summary}).

There are two common approaches to perform DSSE in unobservable systems: providing pseudomeasurements at unobserved power injections \cite{Primadianto}, resulting in a pseudo-observable system, or solving the underdetermined state estimation problem. In this work, we consider the second approach, as we have limited historical data on which to base pseudomeasurement estimates, and because there are non-trivial observable islands within the state estimation problem. (Pseudomeasurements would act to bias the solution within those observable islands, and so may lead to estimation accuracy being reduced when considering nominal DSSE accuracy.)

\begin{figure}
\centering
\subfloat[With nominal MV/LV tap positions]{\includegraphics[width=0.45\textwidth]{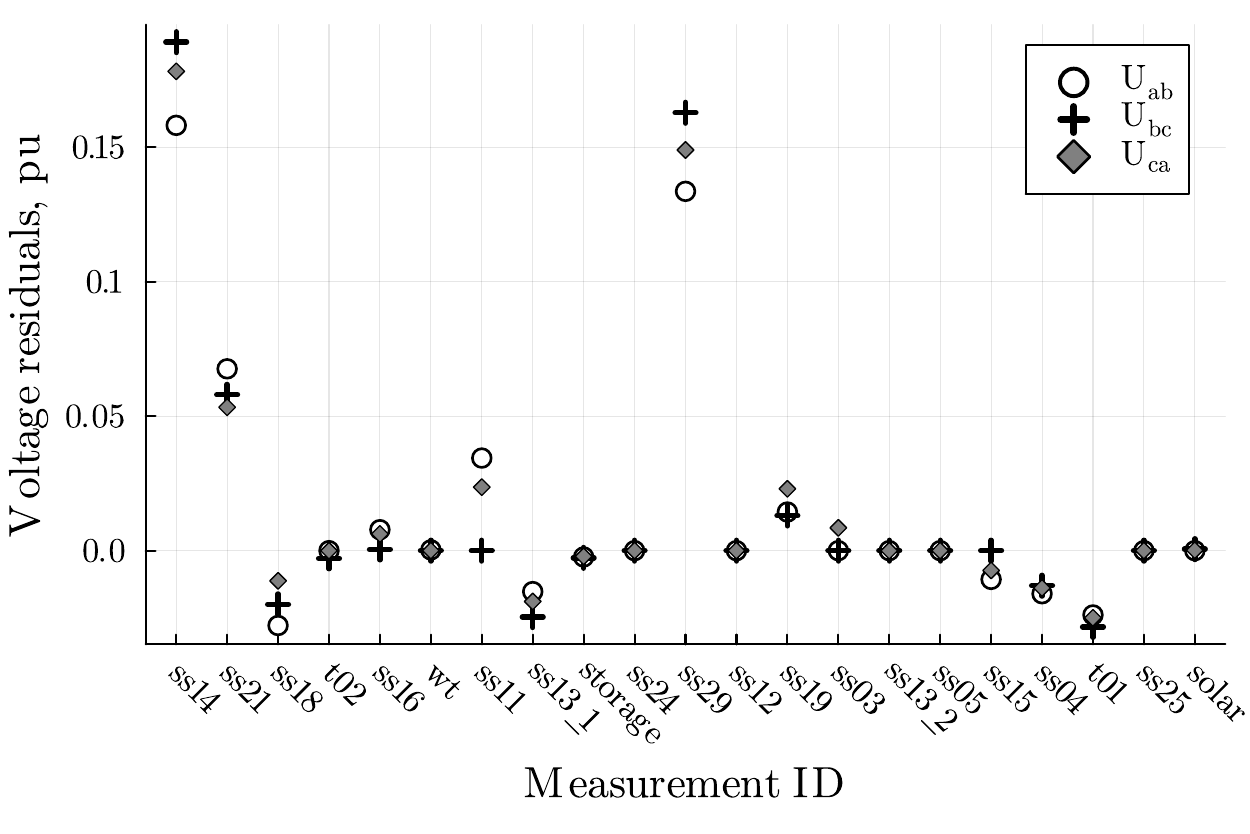}\label{ff:mvlv_nominal_dsse}}
\\
\subfloat[With updated MV/LV tap positions]{\includegraphics[width=0.45\textwidth]{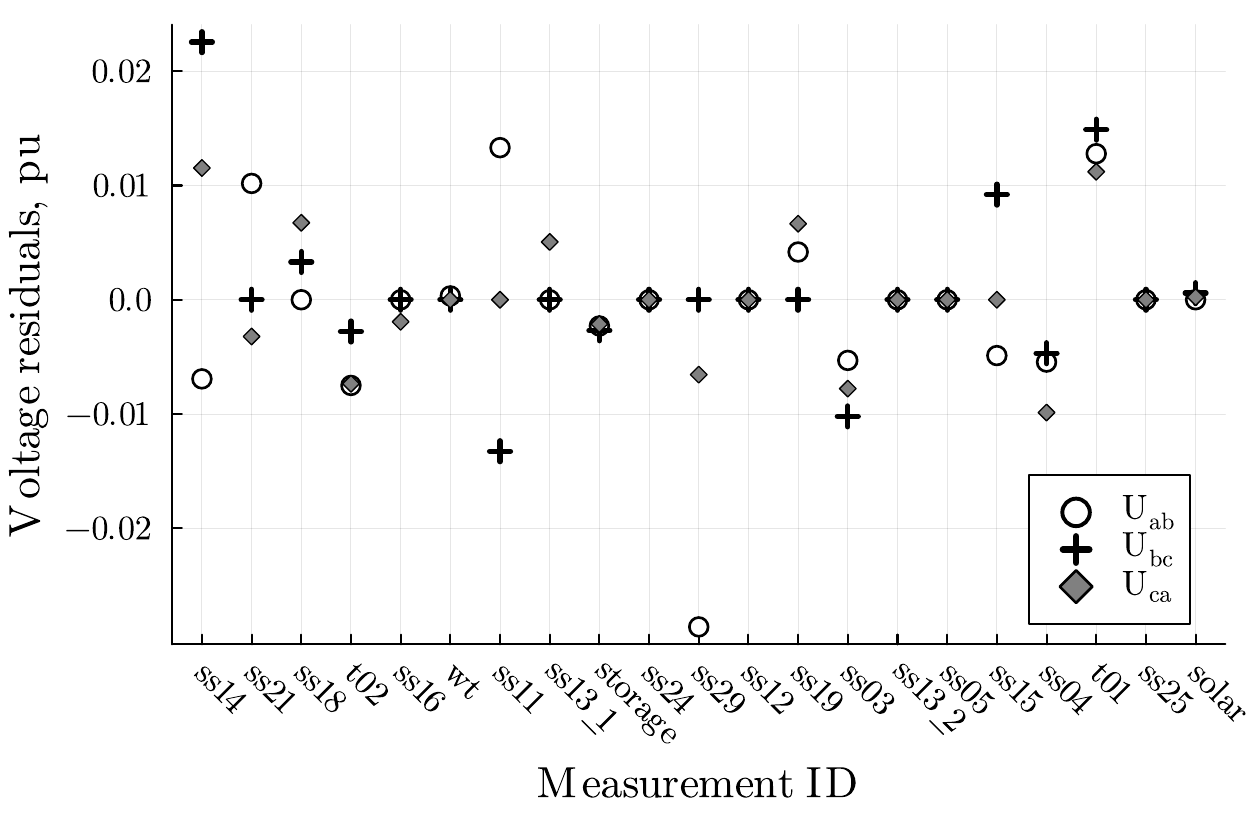}\label{ff:mvlv_updated_dsse}}
\caption{Line voltage residuals following DSSE}
\label{f:residuals_with_power_per_phase}
\end{figure}

The full DSSE formulation is given in the Appendix, and is a non-convex quadratically constrained quadratic programming (QCQP) problem. Results for one timestep are visualized in Fig.~\ref{f:residuals_with_power_per_phase}, both with the nominal and updated MV/LV tap positions (as described in Section~\ref{ss:tap_settings}). These figures plot the residual, calculated as the difference between the measured line voltage and the estimated voltage as calculated via DSSE.

Whilst the true value of state variables cannot be known (due to inevitable measurement noise), approximate bounds can be inferred from meter tolerances (Table~\ref{t:relative_errors_summary}). The results from DSSE for the model with nominal MV/LV tap positions (Fig.~\ref{ff:mvlv_nominal_dsse}) indicate that the measurement and network models are inconsistent. A number of buses are completely unobservable, as can be identified due to their residual having a value that is numerically zero (seven buses have all three voltage residuals less than $ 6 \times 10^{-5}$). These buses have numerical value zero, as in those under-observed subsystems the state estimator will have sufficient degrees of freedom to assign measurement-variable pairs as having identical values whilst still maintaining consistency of the power flow equations.

The output of DSSE following the tap position update is plotted in  Fig.~\ref{ff:mvlv_updated_dsse}, with the residuals having decreased significantly as compared to the nominal MV/LV tap positions. It can be observed that changing the tap positions did not result in a change in residuals for unobservable nodes (as expected). The residuals are still too variable to consider a good DSSE output. Nevertheless, they confirm in a statistical sense that the updated network model improves the Digital Twin's data quality. 

\section{Case Study: Using the Digital Twin for Estimating Benefits of a Dynamic Export Limit}\label{s:curtailment_cases}

The SEND network is subject to a static maximum export limit, set by the utility as a result of voltage congestion, following the UK standard G100 \cite{ena2023g100}. To meet this standard, the SEND energy management system automatically reduces solar power output by several MW if the power threshold is breached for a set time period. When this power reduction occurs, the energy management system acts to increase demand from electric boilers and other flexible demand sources, then the solar export ramps back towards the available power export. However, in general, the maximum solar generation is large in comparison to the energy that can be absorbed by the flexible loads and electric boilers.

In this section, we use the SEND Digital Twin to consider a potential solution to address these high curtailment volumes. In particular, a dynamic export limit is considered, with voltage control also provided by setting a non-unity power factor for the solar generator. In Section~\ref{ss:curtailment_mitigation} we explain the physical mechanism by which this approach would reduce curtailment, with the role of the Digital Twin in calculating the potential benefits of this approach described in Section~\ref{ss:curtailment_model}. The analysis is based on a synthetic test case, with results presented for September 17th (an exemplar day with significant curtailment) in Section~\ref{ss:results}.

\subsection{Reducing Curtailment with a Dynamic Export Limit and Reactive Power Control}\label{ss:curtailment_mitigation}

Changing the SEND connection agreement to have a dynamic export limit would mean that exports will only be limited if relevant state parameters have reached their operational limits at a given time, rather than having only been determined for a conservative, worst-case scenario (as is the case for the existing static export limit). As the SEND system is the only customer which exports power on its feeder in the network (Fig.~\ref{f:mv_system}), it is assumed that there will not be any further voltage rise downstream of it. Therefore, this voltage-based dynamic export limit is equivalent to ensuring that voltage magnitudes do not exceed their limits at the SEND PCC or within its network. As can be seen in Fig.~\ref{f:pltSendMvVoltages} and Fig.~\ref{f:pltSendLvVoltages}, measured voltages indicate significant headroom during many hours of the day, albeit with slight overvoltages during late morning and afternoon hours.

Given the fact that voltage congestion is more likely during periods of high solar output, it is proposed that the power factor of the solar inverter is also adjusted to reduce the impact of generating active power. The benefit of changing generator power factor can be illustrated by considering how a reactive power injection $\Qslr$ and active power injection $\Pslr$ result in a voltage rise  $\Delta U$ due to positive sequence impedance $Z=R+\jmath X$ as (approximately) \cite{turitsyn2011options}
\begin{equation}\label{e:deltaV_approx}
    \Delta U \approx \Pslr R + \Qslr X\,.
\end{equation}
Reducing the power factor to draw $ \Qslr $ (i.e., an increasingly negative $\Qslr$) results in an increased active power $\Pslr$ that can be injected by the solar plant for the same voltage rise. For example, the solar injection that results in a 1\% voltage rise $\PslrRate$ can be calculated from \eqref{e:deltaV_approx} considering the generator power factor $ \PFslr $ as
\begin{align}\label{e:Pslr_1}
   \PslrRate &= \dfrac{1\%}{R - X\cPF}\,, \\
    \cPF &= \dfrac{\sqrt{1-(\PFslr)^{2}}}{\PFslr}\,. \label{e:Pslr_2}
\end{align}
Although \eqref{e:deltaV_approx} is only an approximation, in practise \eqref{e:Pslr_1} provides good accuracy (as as considered in Section~\ref{ss:curtailment_model}), with estimates of $R$ and $X$ determined via power flow linearization techniques.

As per common practise (e.g., \cite{ena2023g100}), it is assumed that operational uncertainty is accounted for in the dynamic export limit control schemes, given the safety-critical function of the controller. Thus, when voltages approach their limits, there will be some curtailment $\Psf$ due to the tolerance with respect to voltage estimates  $\epsilon$ (in \%) as
\begin{equation}\label{e:safety}
     \Psf = \epsilon \PslrRate\,,
 \end{equation}
where $\PslrRate$ is the power injection-voltage rate \eqref{e:Pslr_1}. This means that techniques that reduce $\epsilon $ (e.g., through sensor fusion in DSSE) can reduce the curtailment that would be calculated using raw measurements directly. If only a single voltage measurement is available then the tolerance $\epsilon$ is identical to its operational uncertainty $\epsU$ (as reported in Table~\ref{t:relative_errors_summary}).

\subsection{Using the Digital Twin for Network and Curtailment Modelling}\label{ss:curtailment_model}

There could be a large outlay for SEND to request a dynamic export limit, due to complex network studies for the utility and any additional systems that may be required (e.g., development of utility-side DSSE \cite{ena2012er126}). Given that there is a growing database of historic congestion and curtailment events, the Digital Twin is an ideal candidate to calculate the estimated benefit. With respect to Fig.~\ref{f:digital_twin_workflow}, SEND system planners can use data from the data historian to consider curtailment volumes, use DSSE to provide the best estimate of the historic system state, then use distribution system simulation to consider a counterfactual with a dynamic export limit and voltage control enabled. In turn, the reduced curtailment volumes can be used to calculate additional revenue for the university and reduction in carbon emissions.

As shown in Section~\ref{s:towards_digital_twin}, the real world network model and measurement data are not fully consistent, and so a synthetic test case is used for this case study. A heuristic process combines the network model with power and voltage measurements to create an unbalanced system with the slack voltage and load and generation powers all specified (including allocating load at nodes without measurements). Specifically, for nodes with measured real and reactive powers, these powers were equally split amongst phases. The system residual power (the difference between power injected and total power across measured substations) is then allocated among loads and phases accordingly, with those individual allocated powers uniformly distributed between 0\% and 200\% of the mean per-load average. To reduce computational burden, every fourth measurement is used (i.e., at a two minute temporal resolution).

Once those loads have been allocated, synthetic voltage magnitudes and power measurements are then determined by running a power flow and adding white noise to those synthetic measurements according to the tolerances given in Table~\ref{t:relative_errors_summary} (with standard deviation one third of the total error). To avoid the challenges described in Section~\ref{ss:dsse_and_observability}, the synthetic measurements that are captured consist of phase voltages and per-phase powers at all load and generation buses. Under these conditions, the network is observable and the DSSE can be used reliably.

The amount of solar curtailment at a given time instance is not estimated within the data historian DEOP. Physics-based curtailment estimation is a complex task (requiring, for example, panel temperature, direct and diffuse irradiance, shading, and solar panel electrical parameters). For the purposes of this work, a heuristic approach is therefore used to estimate curtailment. The GB-wide solar profile from \cite{wilson2021electrical} is normalised, scaled and offset according to the installed capacity at the solar generator, to estimate the potential generation during periods of known curtailment. The curtailment is then estimated as the difference between the potential and measured power output (with any spurious negative values assigned a value of zero). The resulting profile was validated by inspection for the date considered.

\subsubsection{Implementation Approach}\label{sss:linearization}
A linear power flow formulation is used to calculate changes to the system voltage magnitudes in SEND $\Vsend$ as compared to the estimated voltage magnitudes from DSSE $\Vtwin$ due to real and reactive injections at the solar generation $\Pslr,\,\Qslr$,
\begin{equation}\label{e:curtail_vcalc}
\Vsend = M\begin{bmatrix}
    \Pslr\\
    \Qslr
\end{bmatrix}
+ \Vtwin \,,
\end{equation}
where $M\in \mathbb{R}^{3n\times 2}$ is the linear sensitivity matrix for a system of $ n $ three phase buses. This approach was chosen as it is conceptually simple, avoiding a need for iterative power flow calculations. The linear matrix $ M $ was created using the approach outlined in \cite{bernstein2018load} and was found to have very good accuracy, with the relative error in the 2-norm of voltage changes found to be between 1\% and 9.7\% or less for per-phase real or reactive injections of 200 kVA at any node in the network.

The maximum possible solar injection $\PslrMax$ for a given power factor will happen at the smallest solar injection $\Pslr$ for which any bus reaches the upper voltage limit $U^{+}$. This can be calculated from \eqref{e:curtail_vcalc} as
\begin{equation}\label{e:Psrl_calc}
\PslrMax = \min (U^{+} - \Vtwin)/_{e}\left (M\begin{bmatrix}
    1\\
    - \cPF
\end{bmatrix}
\right)
\end{equation}
where $/_{e}$ denotes elementwise division and coefficient $\cPF$ maps real to reactive powers for a given fixed power factor, as in \eqref{e:Pslr_2}.

\subsection{Results}\label{ss:results}

\begin{figure}
\centering
\includegraphics[width=0.48\textwidth]{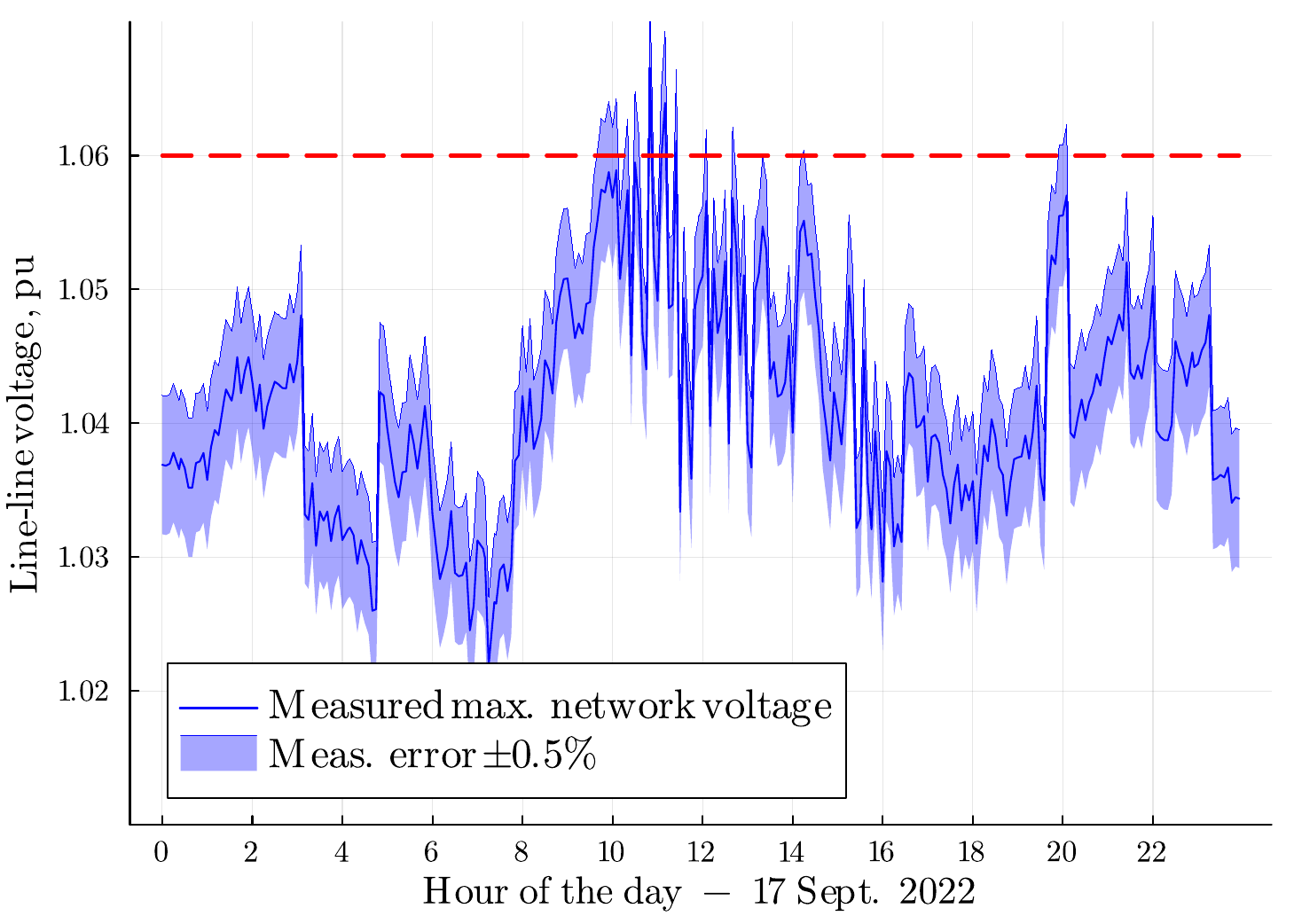}
\caption{Maximum network voltage, including measurement error bars.}
\label{f:pltInjectionExample_volts_fill}
\end{figure}

The maximum (synthetic) MV measured voltage in the SEND system for 17th September is shown in Fig.~\ref{f:pltInjectionExample_volts_fill} alongside the tolerance $\epsU $ of $\pm$ 0.5\%. It can be seen that for short periods, there are over-voltages. A dynamic export limit would avoid these issues (although, it is noted that power quality standards typically allow voltages to stray outside of nominal bounds by a small amount, so long as this is only for a short duration).

\begin{figure}\centering
\includegraphics[width=0.44\textwidth]{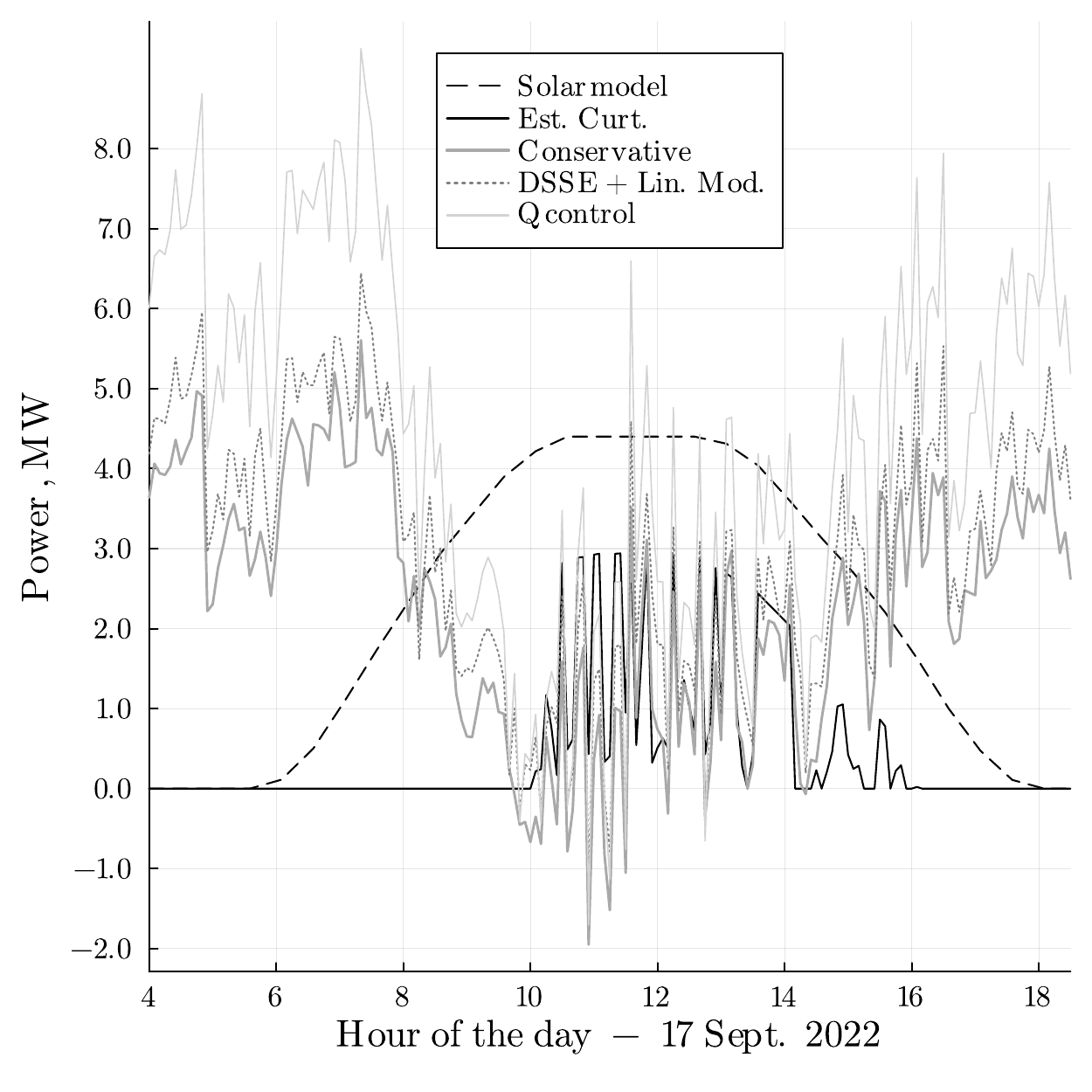}
\caption{The solar generation model (`Solar model'), modelled curtailment (`Est. Curt.'), and additional potential solar generation $\PslrMax$ for three dynamic curtailment schemes.}
\label{f:curtailm}
\end{figure}

We use \eqref{e:Psrl_calc} to calculate the additional solar generation $\PslrMax$ that could be obtained with three different control schemes, instead of static curtailment. Results are shown in Fig.~\ref{f:curtailm}.
\begin{itemize}
    \item A first scheme considers a generator exporting with unity power factor (labelled `DSSE + Lin. Mod.'). This model illustrates the additional benefit of moving to a dynamic limit without additional voltage control. It is assumed that the DSSE process means that the operational uncertainty $\epsilon $ can be assumed to be small (e.g., that $\epsilon \ll \epsU$).
    \item A second scheme uses a power factor of 0.9 (labelled `Q Control'). This scheme is used to illustrate the additional generation possible due to the addition of an example voltage control scheme, also assuming that operational uncertainty $\epsilon$ in DSSE outputs is small.
    \item Finally, a third dynamic scheme considers unity power factor generation, but considers the impact of operational uncertainty $\epsilon $ on the additional generation (labelled `Conservative'). This uses as a voltage for calculations the smaller of the raw measurements and DSSE output plus the operational tolerance $\epsU$ of 0.5\%.
\end{itemize}
The `Conservative' model results in the smallest additional generation $\PslrMax$, and the `Q Control' approach the largest. In addition, on this figure the estimated total solar and estimated curtailment are plotted.

From Fig.~\ref{f:curtailm}, it can be seen that curtailment takes up a substantial fraction of the potential solar generation on this day. Nevertheless, the utilization of the network could clearly be increased substantially--in the hours before 8~am, more than 2~MW of additional power could be generated from the site for all dynamic export schemes, and after 2~pm, there is sufficient capacity even for the conservative scheme to mitigate all present curtailment. There are some negative values of $\PslrMax$ in the late morning period as the measured maximum voltage is greater than 1.06~pu for a small number of time periods.

Note that the differences between the three curtailment reduction schemes can be explained succinctly by considering the voltage-curtailment rate $\PslrRate$ and curtailment safety factor $\Psf$. These are calculated with respect to voltage rise at solar bus in Table~\ref{t:power_factor_voltage}. The power injection-voltage sensitivity $\PslrRate$ increases the power that can be injected by 44\%, resulting in `Q Control' being (close to) a factor of 1.44 times that of `DSSE + Lin. Mod.', as can be observed by-eye. The curtailment safety factor $\Psf $ has value of 0.82~MW, and so the `Conservative' scheme has a constant factor close to this value less than can be exported as compared to `DSSE + Lin. Mod.'.

\begin{table}
\caption{Power injection-voltage sensitivity $\PslrRate$ (from \eqref{e:Pslr_1}) and corresponding safety factor $\Psf$ (from \eqref{e:safety}) for tolerance $\epsilon=0.5$\%.}
\centering
\begin{tabular}{lll}
\toprule
\tcell{Power\\factor} & \tcell{Sensitivity\\$\PslrRate$, MW} & \tcell{Safety factor\\$\Psf$, MW}\\
\midrule
1.0 & 1.640  & 0.820 \\
0.9 & 2.355 & 1.178 \\
\bottomrule
\end{tabular}

    \label{t:power_factor_voltage}
\end{table}

\begin{figure}\centering
\includegraphics[width=0.48\textwidth]{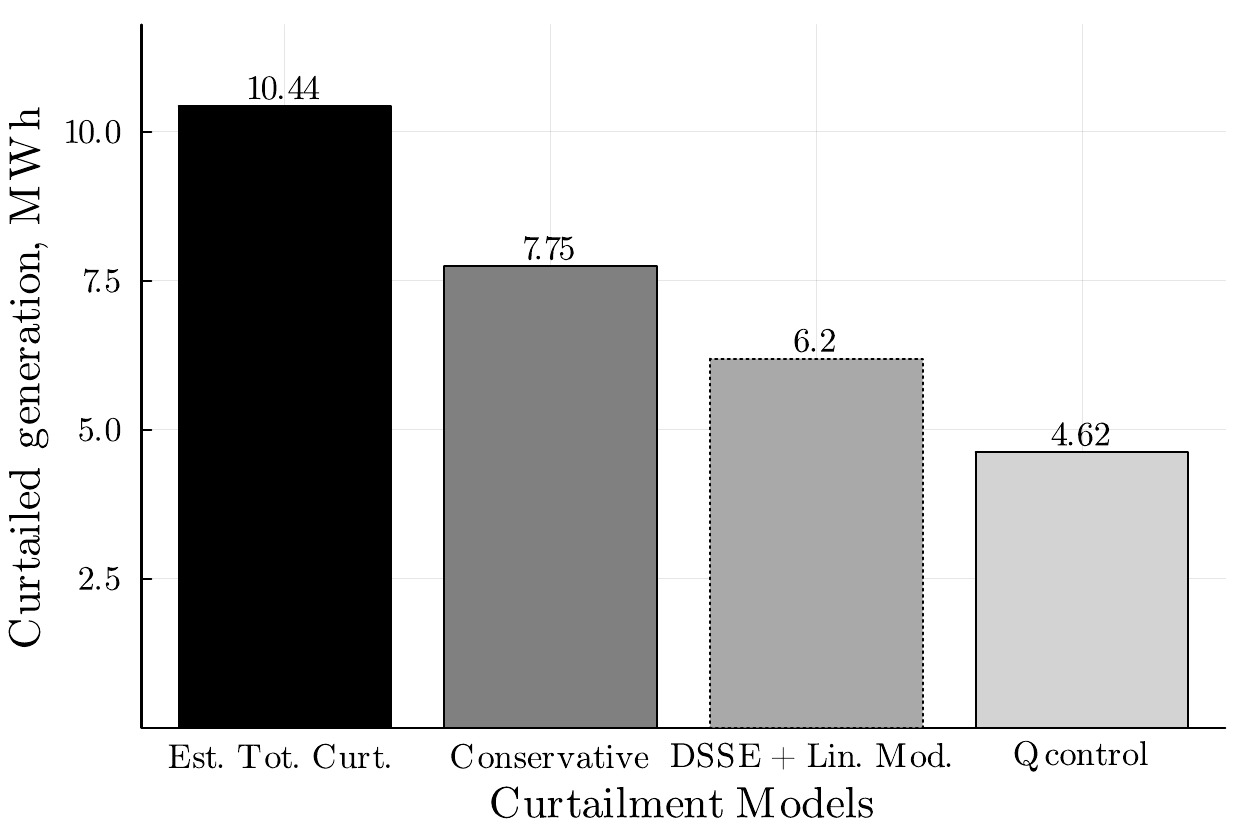}
\caption{Bar chart showing the total estimated curtailment against three control approaches that utilise a dynamic export limit, as compared to the existing static export limit.}
\label{f:pltInjectionExample_curt_bar}
\end{figure}

Fig.~\ref{f:pltInjectionExample_curt_bar} plots the estimated energy curtailed for the three models. It can be seen from this figure that all methods can reduce the curtailment on this day by a significant amount. The approach based on unity power factor control results in a reduction in curtailment of 4.2~MWh, although if the utility insists on conservative state estimation, this reduces the output by 2.69~MWh. In contrast, the `Q Control' model reduces curtailment volumes by 5.82~MWh. For this latter case, an energy price of \$100/MWh and grid carbon intensity of 400~kgCO2e/MWh implies potential to increase revenues by \$582 and reduce emissions by 2.33 TCO2e for this day.

\section{Discussion: Towards Digital Twins}\label{s:next_test_case}

The development of the network models and collection of data for the proposed system has resulted in a `synthetic' Digital Twin of the SEND distribution system, with more work required to develop a self-consistent, fully functioning twin. Much of the software and hardware implementation now exists for the system, but there are the outstanding issues which highlight general challenges that distribution system Digital Twins will face.

\paragraph{Defining the Scope of the Digital Twin Framework} The Digital Twin framework proposed in Section~\ref{s:framework} may need to be adapted to account for additional scope depending on its application. For example, the curtailment model was considered a component of the `data historian' block; however, in applications with many customers (e.g., in the setting of domestic solar PV dynamic operating limits), the forecasting of customer-side DERs becomes a substantial task in itself, with output depending on solar PV capacities, orientation, and shading. Closer connections with other energy sectors, such as with electric vehicles or heating systems, would also require substantial modelling efforts to achieve adequate performance.

\paragraph{Automated Network Model Identification and Correction} As shown in Fig.~\ref{f:digital_twin_workflow}, the network model is influenced by DSSE and network simulation, as the network model must be validated for real-world applications. It has been proposed that DSSE can also be augmented to include the identification of network parameters directly, particularly with respect to tap settings \cite{Nanchian2017,Korres2004}. The approach of these works is to consider how residuals over a number of time periods can be minimized, and so tap settings are reported jointly with the system state. Given the requirement for power meters to be reconfigured to allow reasonable state estimation, this was not considered, but it is noted that these approaches have been shown to be very effective in real-world applications \cite{vanin2022role}.

\paragraph{Maturation of Partially-Observable DSSE Methodologies} In traditional power system applications, the main benefit of state estimation is to provide a systematic approach for recovering the system state using a self-consistent physical system model, and to do so with the highest possible accuracy. In distribution systems, obtaining full observability for all buses and line flows is typically not cost-effective. In the physical DSSE calculations described in Section~\ref{ss:dsse_and_observability}, our approach has been to consider only the estimation of the state via an underdetermined formulation, and so the state is returned only as a non-null value in an observable subset of nodes. However, in a real distribution system, it may be the case that the operator does need an estimate of the full system state. To address this, either the network operator can install further measurement devices, or use pseudomeasurements. However, without good historic data on which to base the pseudomeasurements, the uncertainty of estimated state could be large\footnote{We note that estimating good pseudomeasurements for individual loads, e.g., households, is particularly challenging. This task may be easier for aggregated loads.}. Therefore, even though the state estimator would technically be observable from a mathematical perspective, the true uncertainty can be so large as to be operationally meaningless. Therefore, where pseudomeasurements are necessary, efforts should be made to ensure these are as accurate as possible (e.g., by aggregating lower fidelity smart meter measurements).

Furthermore, DSSE methods tailored to consider partially observable systems could be implemented. In particular, developing solutions to automatically identify observable sub-areas appears a promising research direction. Within those sub-areas reliable state and parameter estimation is possible.

\paragraph{Development of Test Systems for Digital Twins and DSSE} 
We note that other publicly available distribution system test cases (both real-world and synthetic) provide power flows at loads and generators, but not voltage measurements, and so are too simplistic for Digital Twin applications. This is because state estimation and parameter identification tasks require voltages to ensure the system is over-determined, and to capture the complexity that is associated with unknown errors in network models and measurement uncertainty. Given this lack of test cases, authors must make heuristic assumptions as to the quality of metered data, meter placement, and typically assume a known and perfect network model. Additionally, in those cases, noise is added to power flow results, yielding unreproducible residuals (unless the noise is provided together with the results). Furthermore, the structure of meter errors may be very complex--measurement errors can drift with time, leading to correlated errors. DSSE problems never have a known ground-truth, and so real-world validation methods have a different flavour to the results output from synthetic test cases. Finally, a future test system could also consider LV systems more comprehensively, using GIS information and consumer smart meter data, as DSSE in LV networks is typically heavily reliant on those measurements.

\section{Conclusions}\label{s:conclusions}

Digital Twins for distribution systems are set to become a widespread architecture as energy system digitalization strategies reach fruition in coming years. They promise to enable new operational and planning approaches, with potential to maximise the value of data from newly commissioned measurement devices in a self-consistent representation of the physical energy network. In this work, we have proposed a Digital Twin framework, then implemented this for the SEND demonstration campus at Keele University, UK. As a publicly-funded demonstrator project, it was possible to provide open network models, measurement data and algorithms for exploring non-idealities of real-world Digital Twin components. This contrasts with distribution system test systems today, which do not consider where power meters might realistically be installed, their tolerance, or in non-trivial network modelling errors, each of which can have substantial effects on subsequent analysis.

State estimation has been proposed as an integral part of the Digital Twin architecture. The SEND measurement data and network model underlines the necessity of both correctly configured meters and a network model with a very good accuracy to enable an adequate state estimation output. We hope that other researchers can use the model and data for development and validation of new algorithms to address the challenges seen in this project. For example, future work could include integration of DSSE with model identification functionality to address challenges seen in this work around identifying tap settings of MV/LV transformers.

The Digital Twin has been used to explore the potential for dynamic export limits and voltage control to minimize solar curtailment. However, there is huge potential for a range of further applications both in microgrid and grid-connected systems, such as the scheduling of virtual power plant assets, developing consumer dynamic operating envelopes, or preventive and corrective actions during stress events. We conclude that distribution system Digital Twins could be a cornerstone technology for network operators, and that utilities, academia and industry must work together to develop these as a holistic and flexible tool for the wide range of distribution grid characteristics seen around the world.

\section*{Appendix: DSSE Formulation}\label{appendix_formulation2}

In contrast to conventional DSSE, which is typically based on non-linear least squares minimization solved with Gauss-Newton methods, the DSSE algorithms used in this work are based on (equivalent) optimal power flow-based formulations, with the general approach described in~\cite{vanin2022framework}. As described in Section~\ref{ss:dsse_and_observability}-1), 
the main advantage of this approach is that it is easily customizable and can solve under-determined DSSE problems conveniently. For completeness, we reproduce the full formulation here, noting that with respect to the original implementation \eqref{eq:line-to-phase} is added for the SEND case. The DSSE problem is then solved using Ipopt \cite{ipopt}.   

DSSE is performed in the AC-rectangular (ACR) variable space, with phase voltage phasors $\mathbf{U}\in \Cbb{n\times 3}$ and lifted power variables $\mathbf{S}\in \Cbb{n\times 3\times 3}$ represented within the optimization in real and imaginary components,
\begin{align} 
 \mathbf{U}_{i} = \mathbf{U}_{i}^{\Re}+j\mathbf{U}_{i}^{\Im} \; \; \; & \forall i \in \mathcal{N}, \\
\hskip 3em \mathbf{S}_{ij} =  \mathbf{P}_{ij}+j\mathbf{Q}_{ij} \; \; \; & \forall (i,j) \in \mathcal{E} \cup \mathcal{E}^\mathrm{R},
\end{align}
where $\mathcal{N}$ is the set of buses, $\mathcal{E},\, \mathcal{E}^\mathrm{R}$ are the sets of branches in forward and reverse orientation, and \textbf{bold} math font indicating a variable which can be indexed only via the first element (e.g., $\mathbf{S}_{ij}$ will return a $3\times 3$ matrix). Loads and generators are defined via
\begin{equation}\label{e:S_demand_gen}
\mathbf{S}_k^g,\,\mathbf{S}_k^d \quad \forall k \in \mathcal{G} \cup \mathcal{L}\,,
\end{equation}
where $\mathcal{G}, \mathcal{L}$ are the sets of generators and loads respectively. 

SEND voltage measurements are line voltage magnitudes $|\Uline|$, which are not nominally included in the ACR variable space. These can therefore be incorporated into the formulation via the equality constraint
\begin{equation}
\begin{split}
|\Uline _{i,\,\mrpq}|^{2} &= \lvert {U}_{i,\,\mrp} - {U}_{i,\,\mrq} \rvert ^{2} \\
& \forall i \mapsto m \in \mathcal{M}^{{\lvert U \rvert}_{\mrpq}} , \forall \mrpq  \in \{ ab, bc, ca\}. \label{eq:line-to-phase}    \end{split}
\end{equation}
As discussed in Section~\ref{s:towards_digital_twin}, the raw data from the power meters return only total three-phase powers, rather than per-phase quantities. These are split according to the heuristic \eqref{eq:power_per_phase_approx}, and can be linked directly to demand or generator variables \eqref{e:S_demand_gen}.

The DSSE objective function is the weighted sum of the least squares of residuals,
\begin{equation}
\mathrm{minimise} \; \; \sum_{m \in \mathcal{M}} \rho_m\,,
\end{equation}
where weighted residuals $\rho_m$ are defined as
\begin{equation}\label{e:residuals}
\rho _m =  \dfrac{\left( x_m - z_m\right )^2}{\sigma _m^2} \quad \forall m \in \mathcal{M}\,.
\end{equation}
Here, $x_m$ is the system variable corresponding to the measured value $z_m$, with $\sigma _m$ representing the confidence on the measurement accuracy. The raw measurements are line voltages and (heuristic) per-phase powers (used in Section~\ref{ss:dsse_and_observability}), where the synthetic measurements used in the case study are phase voltage and per-phase powers for the synthetic case study (used in Section~\ref{ss:results}).

The power flow equations (Kirchhoff's current law and generalized Ohm's law) complete the DSSE model:
\begin{align}
&\mathbf{S}_{ij} = \mathbf{U}_i \mathbf{U}_i^H (\mathbf{Y}_{ij}+\mathbf{Y}^c_{ij})^H - \mathbf{U}_i \mathbf{U}_j^H \mathbf{Y}_{ij}^H \; \; \; \forall (i,j) \in \mathcal{E}, \\
&\mathbf{S}_{ji} = \mathbf{U}_j \mathbf{U}_j^H (\mathbf{Y}_{ij}+\mathbf{Y}^c_{ji})^H - \mathbf{U}_i^H \mathbf{U}_j \mathbf{Y}_{ij}^H \; \; \;  \forall (i,j) \in \mathcal{E}, \\
&\sum_{k \in \mathcal{G}_i} \mathbf{S}_k^g - \sum_{k \in \mathcal{L}_i} \mathbf{S}^d_k = \sum_{(i,j) \in \mathcal{E}_i \cup \mathcal{E}_i^\mathrm{R}} \mathrm{diag}(\mathbf{S}_{ij}) \; \; \forall i \in \mathcal{N}. 
\end{align}
In the above, $(\cdot)^H$ indicates the conjugate transpose, $\mathbf{Y}_{ij}$ represents the series admittance of the branch's $\Pi$-model (see~\cite{PowerModelsDistribution,ClaeysTransformers}), whereas $\mathbf{Y}_{ij}^c$ is the shunt admittance. $\mathcal{G}_i$, $\mathcal{L}_i$ are the sets of generators and loads connected to bus $i$.

\section*{Acknowledgement}

The authors are grateful for extensive support in accessing data and developing the Digital Twin from Ashley Dean, Matthew Dean and Ian Shaw at SEND, Keele University. The authors also thank Dr Frederik Geth from GridQube for helpful discussions on observability and state estimation. The Smart Energy Network Demonstrator project (ref: 32R16P00706) is part-funded through the European Regional Development Fund (ERDF) as part of the England 2014 to 2020 European Structural and Investment Funds (ESIF) Growth Programme. It is also receiving funds from the Department for Business, Energy and Industrial Strategy (BEIS). M. Deakin was supported by the Centre for Postdoctoral Development in Infrastructure, Cities and Energy (C-DICE) programme (funded by the Research England Development Fund) and the Royal Academy of Engineering under the Research Fellowship programme.


\end{document}